\documentclass[11pt,emulateapj,apj]{emulateapj}
\shorttitle{Environmental Dependence of Galaxy Properties}
\shortauthors{Park et al.}

\begin{document}
%\twocolumn[
\title{Environmental Dependence of Properties of Galaxies\\ 
in the Sloan Digital Sky Survey}

\author{Changbom Park\altaffilmark{1}, Yun-Young Choi\altaffilmark{1},
Michael S. Vogeley\altaffilmark{2}, J. Richard Gott III\altaffilmark{3}
and Michael R. Blanton\altaffilmark{4}\\
(For the SDSS Collaboration)}

\begin{abstract}
% MSV: Add a comment about the range of absolute magnitude probed by
% this paper? Results are mostly for galaxies with M<-18.4. 
% We really do not know much about truly faint galaxies.
We investigate the dependence of physical properties of galaxies brighter than 
$M_r = -18.0 + 5{\rm log}h$ in
the Sloan Digital Sky Survey (SDSS) on environment, as measured by
local density using an adaptive smoothing kernel.  We find that
variations of galaxy properties with environment are almost entirely
due to the dependence of morphology and luminosity on environment.
Because galaxy properties depend not only on luminosity 
but also on morphology, it is clear that galaxy properties 
cannot be determined solely by dark halo mass. 
When morphology and luminosity are fixed, other physical properties,
such as color, color-gradient, concentration, size, velocity dispersion, 
and star formation rate,
are nearly independent of local density, without any break or feature.
The only feature is the sharp decrease of the late type fraction above
the critical luminosity of about $M_r = -21.3$ in the morphology
versus luminosity relation.
Weak residual dependences on environment include that of the color of
late types (bluer at lower density) and of the $L$-$\sigma$ relation
of early types (larger dispersion at higher density for bright
galaxies).  The fraction of galaxies with early morphological type is
a monotonically increasing function of local density and luminosity.
The morphology-density-luminosity relation, as measured in this work,
should be a key constraint on galaxy formation models.
We demonstrate that the dependence on environment of the morphology of galaxies
originates from variations in density on effective Gaussian smoothing scales
much smaller than $12 h^{-1}$Mpc. We find that galaxy morphology varies
both with density measured on an effective Gaussian smoothing scale 
of $4.7 h^{-1}$Mpc and with distance to the nearest bright galaxy,
particularly when the distance is about 0.2 $h^{-1}$ Mpc.
We propose as a mechanism that the morphology of galaxies
in galaxy systems is transformed by the tidal force.
\end{abstract}
\altaffiltext{1}{Korea Institute for Advanced Study, Dongdaemun-gu, Seoul 130-722, Korea}
\altaffiltext{2}{Department of Physics, Drexel University, 3141 Chestnut Street,
Philadelphia, PA 19104, USA}
\altaffiltext{3}{Department of Astrophysical Sciences, Peyton Hall, Princeton University, Princeton, NJ 08544-1001, USA}
\altaffiltext{4}{Center for Cosmology and Particle Physics, Department of Physics, New York University, 4 Washington Place, New York, NY 10003, USA}

\section{Introduction}

The origin of the morphology of galaxies is one of the mysteries of
galaxy formation. Recently, there has been a great deal of effort to
understand the origin of morphology by inspecting dependences of
galaxy properties on environment.  This line of study dates back to
the 1930's (e.g. Hubble \& Humason 1931) when it was realized that
clusters were dominated by ellipticals and lenticulars and that
environmental factors played an important role in determining the
morphology of galaxies.  Oemler (1974) found the
morphology-radius relation; the late type galaxy fraction decreases
with radius within a cluster.  This relation was confirmed by Dressler
(1980) who argued that the fraction of morphological types is a
function of local galaxy density.  Postman \& Geller (1984) extended
this morphology-density relation down to the group environment.
Gunn \& Gott (1972) argued that S0 galaxies in great relaxed clusters like
Coma were the result of spirals being stripped of gas by
ram pressure stripping due to hot intra-cluster gas.
Gott \& Thuan (1976) argued that elliptical galaxies were produced
by larger initial density fluctuations which would have higher density
at turn-around and where star formation would be completed before
collapse. Such larger initial density fluctuations would be more
likely in a region that would later turn into a high density environment.

Environmental dependences of galaxy properties other than morphology
have also been studied extensively.  
%For example, it is found from
%radio observations that spiral galaxies become more deficient of HI as
%they are closer to a cluster center (Magri et al. 1988).  
Recently, as
large galaxy redshift surveys like the Two Degree Field Galaxy Redshift Survey 
(2dFGRS; Colless et al. 2001) and the Sloan Digital Sky Survey (SDSS) 
have been completed, extensive studies have been made to accurately measure the
environmental effects on various physical properties of galaxies
(Goto et al. 2003; Balogh et al. 2004a,b; Rojas et al. 2004, 2005; 
Hoyle et al. 2005; 
Blanton et al. 2005a; Croton et al. 2005; 
Tanaka et al. 2005; Weinmann et al. 2006; 
Boselli \& Gavazzi 2006; among many others).  
% MSV: Need to add and explain references here to other work on 
% environmental effects, including Rojas et al. Hoyle et al, Blanton et al. and
% references in those papers
%%%%%%%%%%%%%
Works done by others on similar topics will be discussed in each
section below when the dependence of each of physical properties 
of galaxies on environment is studied. 
The observed
environment-galaxy property relations provide only indirect clues to
the origin of morphology, but they certainly are important constraints on
models for the physics of galaxy formation.

In a companion paper we study the relations among various
physical properties of galaxies in the SDSS spectroscopic sample (Choi et al. 
2006, hereafter Paper I).
In this paper we extend our work to the environmental dependence of
the physical parameters.  We accurately measure the
morphology-density-luminosity relation from a series of volume-limited
samples drawn from the SDSS.  We also use morphological subsets of
early and late type galaxies to measure the dependences of various
physical properties of galaxies on environment once the morphology and
luminosity are fixed.  
By studying the environmental dependence at different smoothing scales,
we address the important question of whether galaxy morphology depends
primarily on the large-scale environment in which the galaxy initially
formed or on a smaller scale environment that may reflect the influence
of later evolutionary effects such as galaxy-galaxy interactions.

\section{Observational Data Set}

\subsection{Sloan Digital Sky Survey Sample}

We use a large-scale structure sample, DR4plus, 
of the SDSS (York et al. 2000; Blanton et al. 2003a; Fukugita et al. 1996;
Gunn et al. 1998, 2006; Hogg et al. 2001; Ivezi\'{c} et al. 2004;
Lupton et al. 2001; Pier et al. 2003; 
Smith et al. 2002; Stoughton et al. 2002; Tucker et al. 2006) 
from the New York University Value-Added Galaxy Catalog 
(NYU-VAGC; Blanton et al. 2005b).
This sample is a subset of the recent SDSS Data Release 5 
(Adelman-McCarthy et al. 2007).
Our major sample of galaxies used here is a subset of the large-scale
structure DR4plus (LSS-DR4plus)
sample referred to as ``{\tt void0},'' which includes the Main galaxies 
(Strauss et al. 2002) with apparent magnitudes
in the range $14.5$ For the approximately 6\% of targeted galaxies 
that lack a measured redshift because of fiber collisions, we
assign the redshift of the nearest neighbor.

Completeness of the SDSS is poor for bright galaxies with $r<14.5$
because of spectroscopic selection criteria (which exclude objects
with very large flux within the fiber aperture) and difficulties of
automatically measuring photometric properties of very extended
sources.  Due to the magnitude limits of the {\tt void0} sample, the full
range of magnitude of the sample is only $\Delta M=3.1$ 
at a given redshift, and
is even smaller than that in the case of a volume-limited sample with
a finite range in redshift.  As described in detail in section 2.1 of 
Paper I, we add
the missing bright galaxies and thereby extend the magnitude range by
using various existing redshift catalogs which we match to the SDSS
data.  In total, 5195 bright galaxies are added to the {\tt void0} samples,
which means our whole sample includes 317533 galaxies within our
angular sample boundaries (see Fig. 1).  Volume-limited samples
derived from the resulting catalog have nearly constant comoving
number density of galaxies in the radial direction for redshifts $z\ge
0.025$. We treat our final sample as effectively having no bright
limit at redshifts greater than $z=0.025$.  More details about this
sample can be found in Paper I.
 
\begin{figure*}
\center
\includegraphics[scale=0.8]{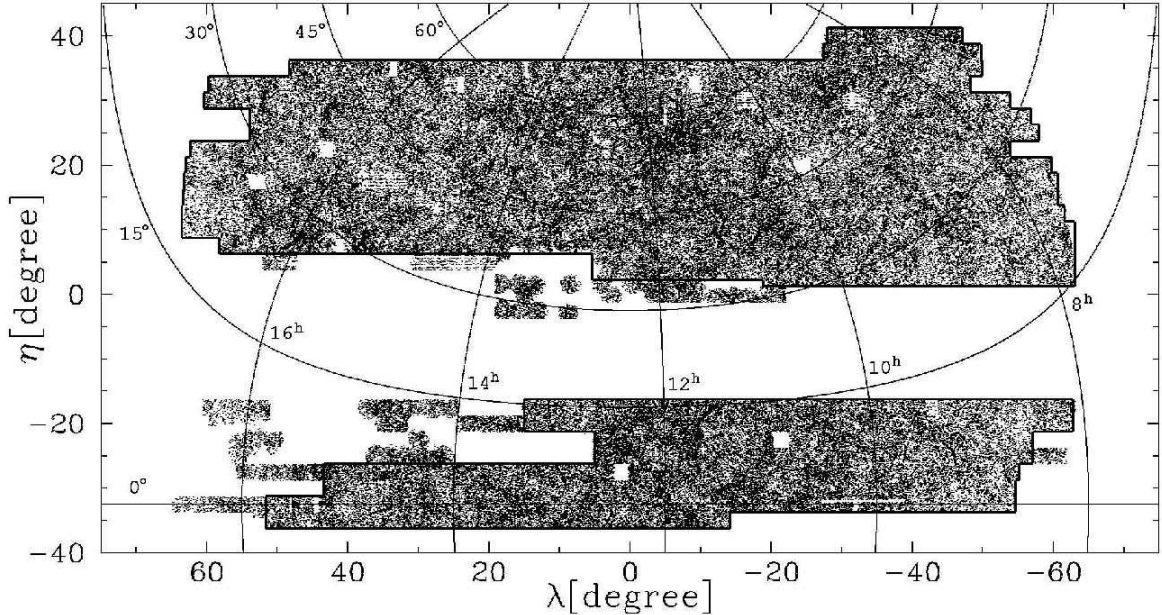}
\caption{Angular definition of the SDSS sample used for our
environmental effect study. Solid lines delineate the boundaries of the analysis
regions in the survey coordinate plane $(\lambda, \eta)$. In addition to
these boundaries the angular selection function in each spherical polygon
defined by spectroscopic tiling is fully taken into account in our analysis.
We name the lower and upper areas regions 1 and 2, respectively.
}
\end{figure*}
 
\subsection{Volume-limited Samples}

To study the effects of environment on galaxy properties it is
advantageous for the observational sample to have the lowest possible
surface-to-volume ratio, to minimize the boundary effects (see Park et
al. [2005] regarding similar effects in topology analysis). For this
reason we trim the DR4plus sample as shown in Figure 1, in which gray
lines delineate our sample boundaries in the survey coordinate plane
[($\lambda, \eta$); Stoughton et al. 2002].  There are also three stripes 
in the Southern
Galactic Cap observed by SDSS that we do not use because of their
narrow angular extent.  These cuts leave approximately $4464$ square
deg$^2$ in the survey region.  Within our sample boundaries, we
account for angular variation of the survey completeness by using 
the angular selection function defined in terms of spherical polygons (Hamilton 
\& Tegmark 2004), which takes into account the incompleteness 
due to mechanical spectrograph constraints, bad spectra, or bright
foreground stars.
We make two arrays of square
pixels of size $0.025^{\circ} \times 0.025^{\circ}$ in the ($\lambda,
\eta$) sky coordinates covering our analysis region, and store the
angular selection function calculated using the MANGLE routine
(Hamilton \& Tegmark 2004).  At the location of each pixel, the
routine calculates the survey completeness in a spherical polygon
formed by the adaptive tiling algorithm (Blanton et al. 2003a) used for
the SDSS spectroscopy. The resulting useful area (with nonzero
selection function) within the analysis regions is 1.362 sr.  In our
study we use only volume-limited samples of galaxies, defined by
absolute magnitude and redshift limits as shown in Figure 3 of Paper I.  
In addition to the volume-limited samples, labeled D1 to D5, which are 
used to study the environmental effects, we also use
the $L_\ast$ galaxy (LS) sample.  The LS sample
is designed to include `typical' $L_\ast$ galaxies ($-20.0\geq
M_{r}\geq -21.0$), and covers the full range of comoving distance, from
$R=74.6$ to 314.0 $h^{-1}$Mpc, in which our analysis is made. All
measurements of local density in this paper use the LS sample to
define the local environment (see section 3).  The definitions for all
samples are summarized in Table 1.

We analyze our samples in the survey coordinate frame, and
the Cartesian coordinates are chosen to be 
\begin{equation}
x = -R \sin\lambda,
y = R \cos\lambda \cos\eta,
z = R \cos\lambda \sin\eta.
\end{equation}
The $y$-axis is at the center of the survey coordinate and radially outward,
the $z$-axis is toward the $(\eta, \lambda) = (90^{\circ}, 0^{\circ})$ direction, 
and the $x$-axis is at the $(0^{\circ}, -90^{\circ})$ direction. 

%%%%%%%%%%%%%%%%%%%% Table 1 %%%%%%%%%%%%%%%%%%%%%%%%%
\begin{deluxetable*}{lccccc}
%\tabletypesize{\footnotesize}
\tabletypesize{\small}
\tablecaption{Volume-limited Samples}
\tablecolumns{6}
\tablewidth{0pt}
\tablehead{
\colhead{Name} &\colhead{Absolute Magnitude} &\colhead{Redshift}&
\colhead{Distance \tablenotemark{a}}&
\colhead{Galaxies$(N_{E}\tablenotemark{b})$}&
\colhead{$\bar d$\tablenotemark{c}}
}
\startdata
D1& $-18.0>M_{\rm r}$& $0.025<z<0.04374$& $74.6<R<129.9$& 20288 (6256)& 3.41\\
D2& $-18.5>M_{\rm r}$& $0.025<z<0.05485$& $74.6<R<162.6$& 32543 (11341)& 3.78\\
D3& $-19.0>M_{\rm r}$& $0.025<z<0.06869$& $74.6<R<203.0$& 49571 (19270)& 4.18\\
D4& $-19.5>M_{\rm r}$& $0.025<z<0.08588$& $74.6<R<252.9$& 74688 (33039)& 4.58\\
D5& $-20.0>M_{\rm r}$& $0.025<z<0.10713$& $74.6<R<314.0$& 80479 (39333)& 5.56\\
LS& $-20.0>M_{\rm r}>-21.0$& $0.025<z<0.10713$& $74.6<R<314.0$& 66176 (30715)& 5.94\\
\enddata
\tablenotetext{a}{Comoving distance in units of $h^{-1}$Mpc}
\tablenotetext{b}{Number of early type galaxies}
\tablenotetext{c}{Mean separation of galaxies in units of $h^{-1}$Mpc}
\end{deluxetable*}
%\end{table}
%%%%%%%%%%%%%%%%%%%%%%%%%%%%%%%%%%%%%%%%%%%%%%%%%%%%%%%%
\subsection{Physical Parameters of Galaxies}

The physical parameters considered in this study are absolute
Petrosian magnitude in the $r$-band $M_r$, morphology, $^{0.1}(u-r)$
color, Petrosian radius, axis ratio, concentration index, color
gradient in $^{0.1}(g-i)$ color, velocity dispersion, and equivalent
width of the H$\alpha$ line.
The $r$-band absolute magnitude $M_r$ is the AB magnitude
converted from SDSS magnitudes. 
To compute colors, we use extinction (Schlegel et al. 1998) and $K$-corrected model magnitudes.
The superscript 0.1 means the rest-frame magnitude $K$-corrected to
redshift of 0.1 (Blanton et al. 2003b). 
This makes galaxies at $z=0.1$ have $K$-correction of $-2.5 \log
(1+0.1)$, independent of their spectral energy distributions (SEDs). 
All of our magnitudes and colors follow this convention,
and the superscript will be subsequently dropped. 
We also drop the $+5{\rm log}h$ term in the absolute magnitude.
We use the luminosity evolution correction of $E(z)=1.6(z-0.1)$
(Tegmark et al. 2004).

We employ the $g$- and $i$-band atlas image and basic photometric parameters
measured by the Princeton/NYU group
\footnote{\url{
http://photo.astro.princeton.edu}} to measure the key physical
parameters of the 317,533 galaxies in our sample.  The $g-i$ color
gradient is given by the color difference between the region with
$R<0.5R_{\rm Pet}$ and the annulus with $0.5R_{\rm Pet}<R<R_{\rm
Pet}$, where $R_{\rm Pet}$ is the Petrosian radius.  We use the
isophotal position angle, the seeing-corrected isophotal axis ratio 
in the $i$-band, and
elliptical annuli in all parameter calculations to take into account
flattening or inclination of galaxies.  The (inverse) concentration
index, $c_{\rm in}$, 
is defined as $R_{50}/R_{90}$ where $R_{50}$ and $R_{90}$ are
the radii from the center of a galaxy containing 50\% and 90\% of the
Petrosian flux in the $i$-band, respectively.  The $g-i$ color
gradient and concentration index are corrected for the effects of
seeing using Sersic model fits of the $g$- and $i$-band images
as described in Paper I.
We choose the concentration
index as a measure of surface brightness profile because the Sersic
indices measured from SDSS images are quite inaccurate except for
large bright galaxies with $r\leq 16$, due to the relatively large
seeing($\sim 1.4''$) of the SDSS.  
If the profile fit is made by avoiding the central part 
of galaxy images, which is much affected by the seeing, 
the seeing effects on the Sersic index can be reduced
but the fit becomes more unreliable due to the smaller fitting
interval and noisier profile. 
In contrast, by definition $c_{\rm in} =R_{50}/R_{90}$ 
is insensitive to the surface brightness profile 
near the galaxy center within radius of
$R_{50}$. Our seeing-corrected concentration index seems 
accurate because it shows no dependence on either redshift or seeing.  

We classify morphological types of galaxies using the prescription of
Park \& Choi (2005). Galaxies are divided into early (ellipticals and 
lenticulars) and late (spirals and irregulars) types based on their locations
in the $u-r$ versus $g-i$ color gradient space and also in the 
concentration index space. 
The resulting morphological classification has completeness and 
reliability reaching 90\%, as claimed by Park \& Choi (2005). 
When photometry is excellent and galaxy images are well-resolved,
more information from surface brightness fluctuations can be added
for morphology classification. But this is certainly not the case
near the faint limit ($r=17.6$) of the SDSS sample we use.
For the volume-limited sample D2 we perform an additional visual
check of the color images of galaxies to correct misclassifications 
by the automated scheme for about $20,000$ blue early types 
(those below the straight line in Fig. 7 of Paper I) and red late 
types (those with $u-r$ color redder than 2.4). The morphological
types of 1.9\% of galaxies, which are often blended
or merged objects, are changed by this procedure.
%This additional visual correction is made to make sure that what we find
%about the relations between galaxy parameters and environment are genuine 
%and, not the
%artifacts due to the small amount of morphology misclassification.

We use Petrosian radius to represent the physical size of galaxies. 
The Petrosian radii in our analysis, calculated by using elliptical annuli, 
are usually larger than those in the 
DR5 catalog, which adopted circular annuli. The angular size is converted 
to a physical size in units of $h^{-1}$kpc from the redshift assuming a 
$\Lambda$CDM universe with $\Omega_{\Lambda}=0.73$ and $\Omega_m =0.27$. 

The stellar velocity dispersion of SDSS galaxies is measured by an
automated pipeline called {\tt IDLSPEC2D} version 5 
(D. J. Schlegel et al. 2007, in
preparation). Galaxy spectra of SDSS galaxies are obtained by optical
fibers with angular radius of $1.5''$.  The finite size of the sampled
light smooths the central velocity dispersion profile. To correct the
central velocity dispersion for this smoothing effects we have adopted
a simple aperture correction formula $\sigma_{\rm corr}=\sigma_{\rm
fiber} (8R_{\rm fiber}/R_0)^{0.04}$ (J{\o}rgensen et al. 1995;
Bernardi et al. 2003a), where $R_0$ is the equivalent circular effective radius
$((b/a)^{1/2}_{\rm deV} r_{\rm deV}$ and $r_{\rm deV}$ is the seeing-corrected effective
angular radius in the $i$ band. 

\section{Local Density Estimation}

To find relations between intrinsic physical properties of galaxies
and their environment we require a well defined and robust measure of
environment.  We seek to use an environmental parameter that is
defined directly from observational data, is continuous, and 
characterizes the full range of galaxy environments, from the most 
massive clusters to voids. The local number density of galaxies 
at a given smoothing scale is such an example and will be 
the environmental parameter in our study.  
Local density gradient and shear can be also used to define
environment, but will be considered in later studies.  Our choice
excludes parameters like the discrete environmental types such as
cluster, group, or field.  Assignment of galaxies to such
environmental types are arbitrary to some extent and can be
constructed based on continuous parameters when desirable.

Local density estimation depends on the shape and size of the smoothing kernel.
We adopt the spline kernel with adaptive smoothing scale, $h_s$ 
to include
a fixed number of galaxies within the kernel weighting.  This adaptive
smoothing kernel is often used in smoothed particle hydrodynamics
simulations.  We use the spline kernel because it is centrally
weighted, unlike TopHAT or cylindrcal, and has a finite tail, unlike the
Gaussian.  The spline kernel was first used to estimate the local galaxy
density by Park et al. (1994) who demonstrated that the
under-dense regions in the Center for Astrophysics (CfA) survey lack bright
galaxies, indicating an environmental dependence of the luminosity
function (LF).  Details of our method to calculate the local density using
the spline kernel are summarized by C. Park et al. (2007, in preparation), who
demonstrate that the smooth density field constructed by the spline
kernel in redshift space recovers the real space density field much
better than the cylindrical kernel at low- and intermediate-density
regions. The local density of galaxies sitting at very high density
regions and with large peculiar velocities (i.e. those appearing at
the tips of fingers-of-god in redshift space) is better measured by
the cylindrical kernel, but the fraction of such galaxies is very
small.

\begin{figure}
\plotone{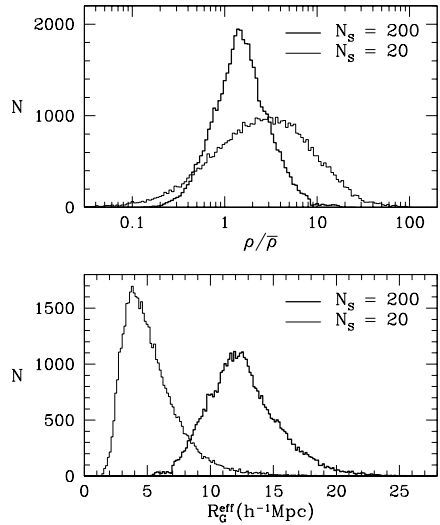}
\caption{Distributions of the local density ({\it top}) and
the effective Gaussian smoothing length estimated at the location 
of galaxies in the D3 sample ({\it bottom}).
The number of $L_\ast$ galaxies included within the spline kernel
are 20 ({\it thin histogram}) and 200 ({\it thick histogram}).
}
\end{figure}

Because this smoothing kernel is adaptive, it has the important
feature that it does not oversmooth dense regions, thereby blurring
the ``walls'' into the ``voids.'' Cosmological simulations clearly show
that the transition from the interior of large cosmic voids, which have nearly
constant density, to the denser structures, defined by filaments and
clusters, is quite abrupt. Maps of the observed
redshift space distribution of galaxies support this picture (see Fig. 3, below).
% MSV: Add ref's to support the previous statement
An adaptive smoothing kernel, such as the one we employ,
better preserves that property
of the galaxy distribution than a kernel with a fixed smoothing
length or with an infinite tail.

% MSV: Do you agree with the following? 
% I'm not sure - couldn't there be small voids in between clusters?
%Do you want this statement here, or  move to Paper III?
On the other hand,
the adaptive nature of the smoothing kernel potentially complicates
the interpretation, because the physical smoothing scale systematically
varies with environment, from a small smoothing scale in clusters to a
very large smoothing scale in voids. 
% MSV: Discuss the advantage of constant signal/noise, or is this in Paper III?
However, the relationship between
environment and smoothing scale is nearly monotonic and, unlike
fixed-scale methods, the adaptive kernel has the advantage that the
S/N for estimating density is more uniform.

% MSV: It might be useful, here or in Paper III, to show how the smoothing
% scale R_S varies. Here are some ideas:
% - Plot of the histogram of R_S for N_S=20 and N_S=200
% - Plot of histogram of \rho/\bar{\rho}
% Most interesting plot I can think of is a scatter plot of
% location of galaxies in R_S versus \rho/\bar{\rho} space.
% The other plots suggested are just projections of that on to one axis.

We define the local density at a position in space as the spline
kernel-weighted number density of $L_\ast$ galaxies surrounding that
position. In every case we use the volume-limited LS sample of $L_\ast$
galaxies with absolute magnitudes $-21.0\le M_r\le -20.0$ for local
density estimation. In this way the local density has a clear meaning,
and it becomes possible to compare the local densities of galaxies in
different volume-limited samples sitting at different radial
distances. The number of $L_\ast$ galaxies required
within the smoothing volume is set to $N_s =20$, which we find is the
smallest number yielding good local density estimates. 
Unless otherwise stated, $N_s =20$; below we also examine results with 
$N_s =200$ when studying larger scales.

When searching for the spline radius containing 
the required number of galaxies, 
the smoothing volume can hit the sample boundaries or
masked regions due to bright stars or galaxies.  In addition to this
trouble, the angular selection function varies across the sky even
within the survey region.  We first check if the location in which the
local density is being measured is within the survey area, and then
ignore the survey geometry and angular selection function to search
for the nearest 20 galaxies in the LS sample in order 
to find the spline radius and
thus the local density.  We then randomly throw 1000 test particles
into the spline radius around the location, and calculate the sum of
normalized weights of the test particles assigned by the spline 
kernel, the angular selection function, and the radial boundaries.
The final local density is given by the first estimate divided by the
sum of weights. Our results use only those local density estimates for
which the correction factor is smaller than 2.

The size of the spline kernel
$h_s$
%'s are 5.0 and $10.8 h^{-1}$Mpc, respectively.  Their corresponding 
can be related with the effective Gaussian smoothing scale
using the fact
that the effective smoothing volumes are ${\pi}h_{s} ^3$ 
and $(2\pi)^{1.5} {R_G}^3$ for spline and Gaussian smoothings.
The result is $R_G = 0.584h_{s}$.
In the case of galaxies in Sample D3, we find 
the median values and 68\% limits of $h_{s}$ are $8.1^{+5.1}_{-2.5}$ 
and $21.0^{+9.0}_{-3.9}$ or the effective $R_{G} \approx 4.7^{+3.0}_{-1.5}$
and $12.3^{+5.3}_{-2.3}$ when $N_s=20$ and 200 are used, respectively.
The corresponding distributions of the local density and the effective Gaussian
smoothing length are shown in Figure 2.
% MSV: these smoothing scales are some effective length. But what can we
% say about variation in the smoothing scale? The variation must be less
% when using a larger N_S.

\section{Environmental Dependence of Physical Parameters}
\subsection{Morphology}

The morphology-luminosity relation has been examined in Paper I.  
In this paper we examine this relation as a function of local
density environment, using our adaptive smoothing to label regions
of space.  The upper panel of Figure 3 shows the distribution of early
(red) and late (blue) type galaxies in the D3 sample in a
wedge defined by $26.25<\eta<36.25$ and $-55.2 < \lambda < 51.6$ in
region 2 (see Fig. 1) of the SDSS.  
% MSV: add reference for the definition of survey coordinates **YY**
The bottom panel shows the same galaxies, 
but distinguished by their absolute magnitudes; 
bright galaxies with $M_r \leq -20$ are red and fainter galaxies with
$M_r > -20$ are blue.  
% MSV: Careful here and through about use of ``bright'' and ``faint.''
% Ordinarily, M>-20 is not considered ``faint.'' Be careful to sayer
% ``fainter'' or ``relatively fainter'' or something like that.

Figure 4 shows distribution of galaxies in a slab with
constant thickness of $20 h^{-1}$Mpc that cuts through region 2. Overlaid
are isodensity contours estimated from the local densities on
a uniform grid through the middle of the slab.  The upper and lower
panels use the local density calculated using the nearest $20$ and $200$
$L_\ast$ galaxies, respectively. This comparison is made to study the
dependence of environmental effects on scale in section 5.  Contours
delineate isodensity regions of $L_\ast$ galaxies.  The thin lines mark
the regions where the local density is $\rho/\bar{\rho} =5$ or 0.1 for
$N_{s} = 20$ and $\rho/{\bar\rho} =0.3$, 2, or 3 for $N_{s} = 200$,
respectively.  The thick curves correspond to the local density equal
to the mean density of the $L_\ast$ galaxies, 
which is measured to be ${\bar \rho}=(5.94 h^{-1}{\rm Mpc})^{-3} =
4.77\times 10^{-3} (h^{-1}{\rm Mpc})^{-3}$.
We clearly see that early types prefer high-density regions, while
late types are more uniformly distributed. Note carefully that the
galaxies plotted are brighter than $M_r = -19.0$ while the density field
is determined by $\sim L_\ast$ galaxies.  Throughout this paper we use
the LS sample to estimate local density, regardless of which
volume-limited sample is being examined.
\begin{figure*}
%\epsscale{1.2}
\plotone{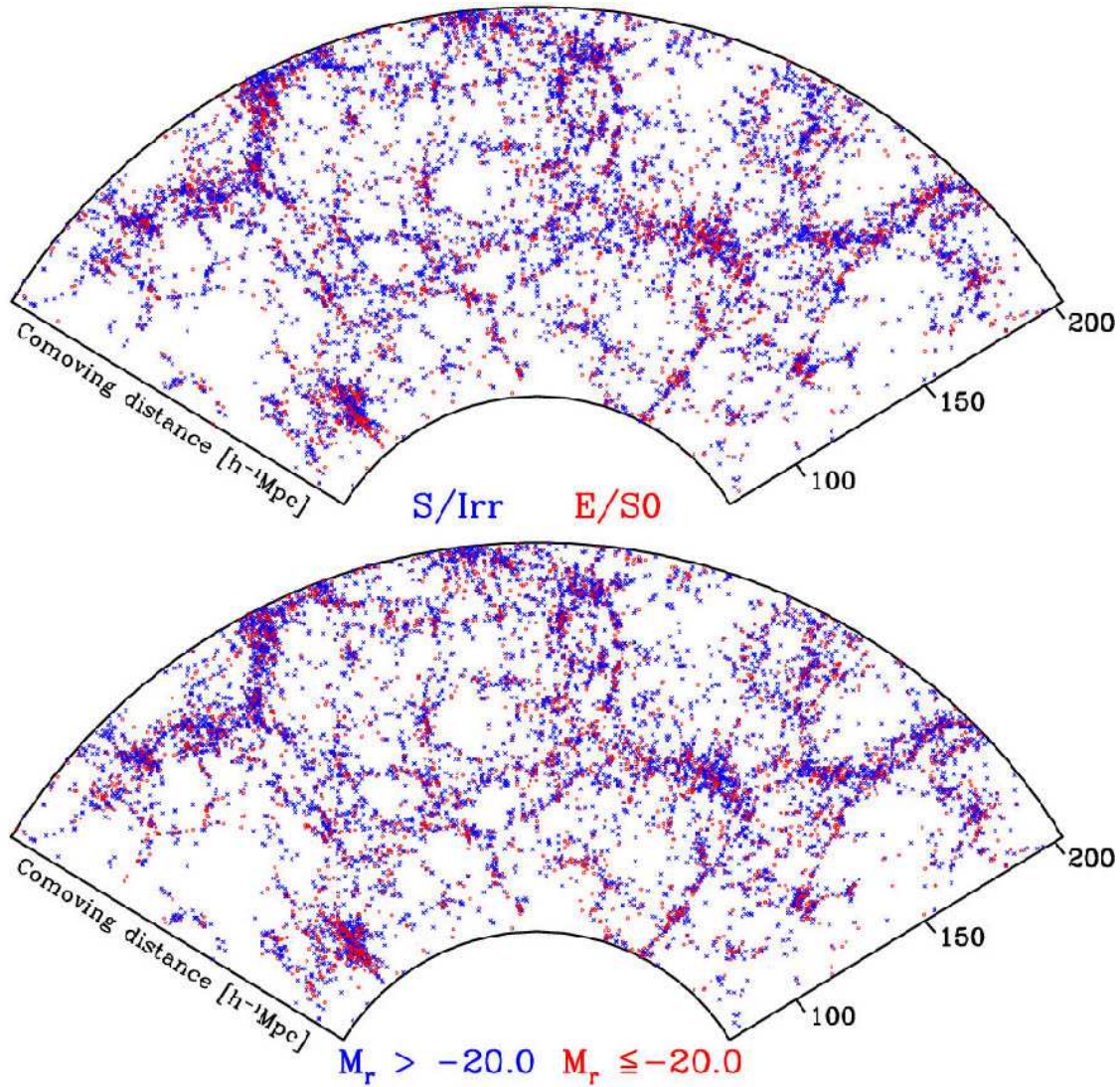}
\caption{{\it Top}: Distribution of early and late morphological types
of galaxies in the D3 sample in a wedge defined by survey coordinate limits of 
$26.25<\eta<36.25$ and $-55.2 < \lambda < 51.6$. Early types are red, 
and late types are blue.
{\it Bottom}: Distribution of galaxies with absolute magnitudes 
$M_r \leq -20$ ({\it red}), and $> -20$ ({\it blue}) in the D3 sample. 
The lower wedge has the same angular definition as the upper one.
Abell clusters A2197 and A2199 are located at the lower left corner of
the wedges.
}
\end{figure*}

\begin{figure*}
%\epsscale{1.2}
\plotone{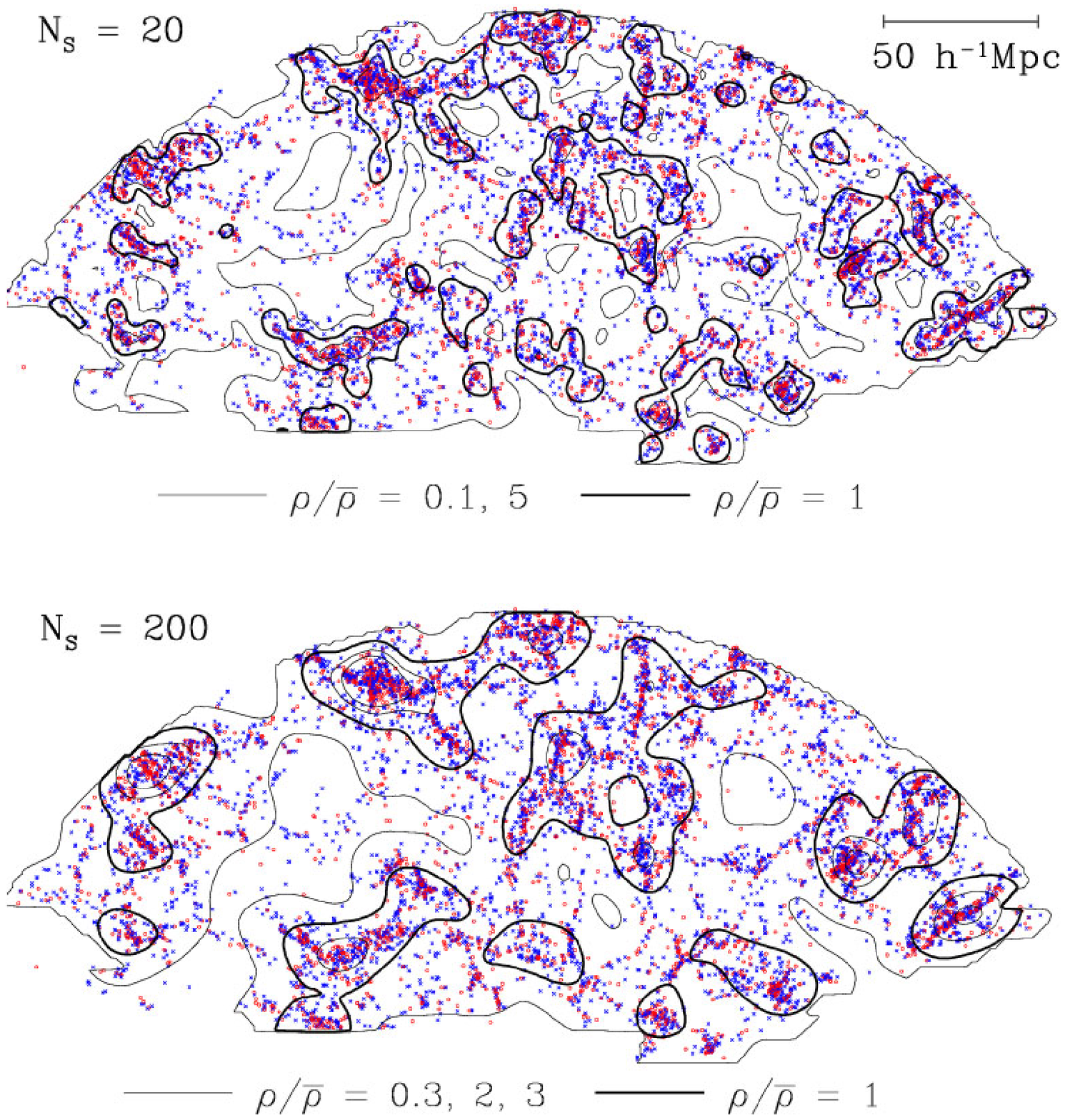}
\caption{Distribution of galaxies in a slab with a constant thickness
of $20 h^{-1}$ Mpc from the $z$-axis coordinate of $35 h^{-1}$ Mpc
to $55 h^{-1}$ Mpc passing through region 2.
Circles are early types, and crosses are late types.
The iso-density contours mark the regions with local density of
$L_\ast$ galaxies of 5, 1 ({\it thick curves}), and 0.1, relative to the mean.
In the upper panel, the smoothing volume includes 20 $L_\ast$ galaxies,
while it is $200$ in the lower panel, where 
the thin lines mark the regions with the local density of 
$\rho/{\bar\rho} =0.3$, 2, or 3.
The mean density of the $L_\ast$ galaxies is 
${\bar \rho}=(5.94 h^{-1}{\rm Mpc})^{-3} =
4.77\times 10^{-3} (h^{-1}{\rm Mpc})^{-3}$.
}
\end{figure*}
 
Figures 5 and 6 present the morphology-density-luminosity relation of 
SDSS galaxies. 
Figure 5 shows the fraction of early type galaxies as a function of 
local density when the absolute magnitude is fixed. 
We first remove late type galaxies in each sample 
whose seeing-corrected $i$-band isophotal axis ratio 
is $b/a<0.6$ in order reduce the internal absorption effect on absolute 
magnitude (cf. Paper I).
\begin{figure}
\epsscale{1.0}
\plotone{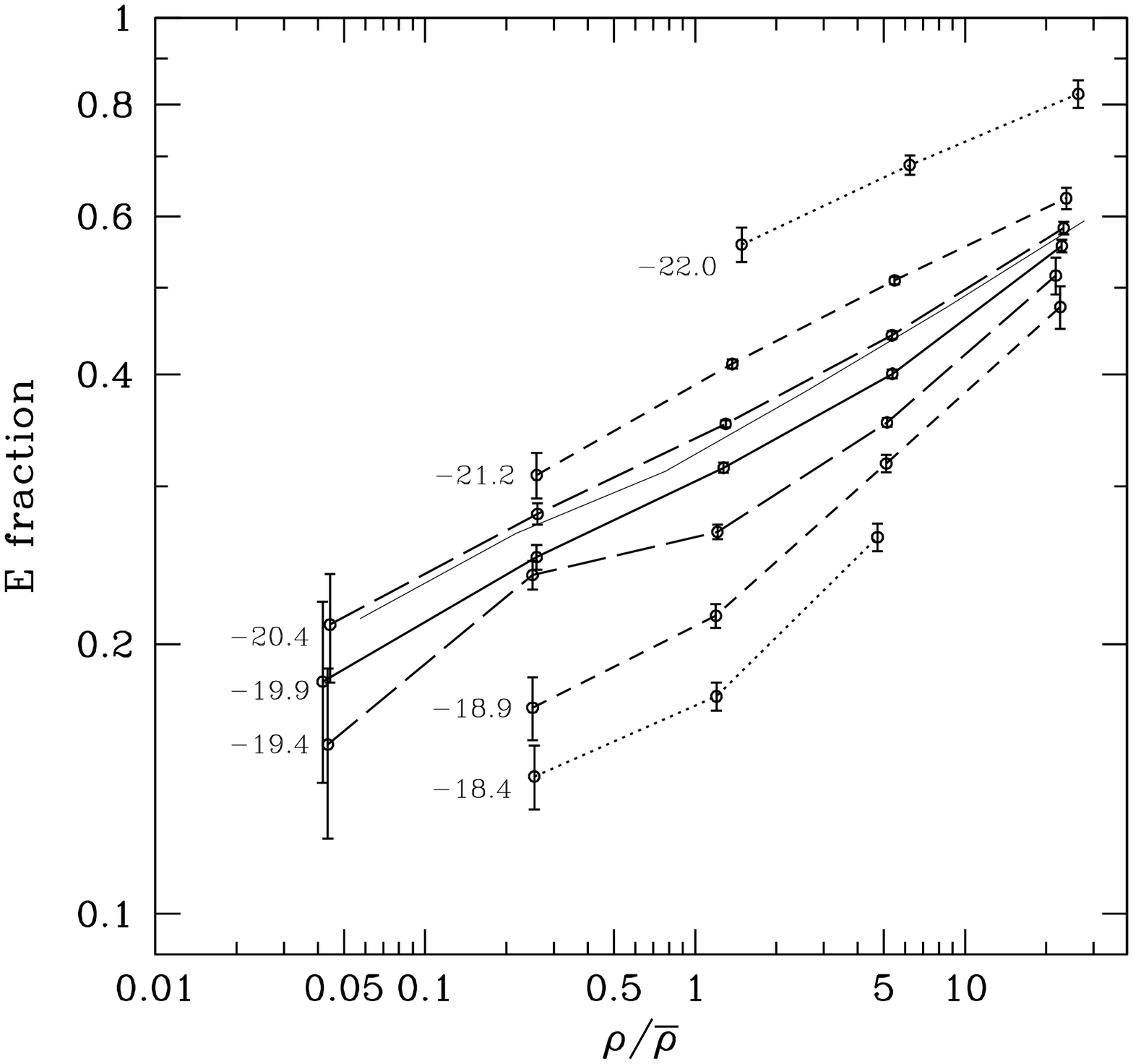}
\caption{Fraction of early type galaxies as a function
of local density at a fixed absolute magnitude (number
assigned to each line). 
The magnitude bins used to draw the thick lines are
$-18.4\pm 0.4, -18.9 \pm 0.4, -19.4\pm 0.4, -19.9\pm0.4,
-20.4\pm0.4, -21.2\pm0.4$, and $-22.0\pm 0.4$, from bottom
to top, respectively. 
The local density bin size is $\Delta {\rm log} \rho = 0.8$.
The thin line is for $M_r=-20.25\pm 0.25$ and
bin size density $\Delta {\rm log}\rho=0.6$.
}
\end{figure} 
We then bin galaxies in the absolute magnitude versus local density space. 
The bin size in this two-dimensional parameter space is 
$\Delta {\rm log} \rho = 0.8$ and $\Delta M_r = 0.8$. 
We measure the fraction of early type galaxies, the median 
local density, and the median absolute magnitude of galaxies belonging 
to each bin, taking into account the fact that we have excluded some 
fraction of late type galaxies. 
When we measure the fraction of early type galaxies, we need to know 
the number of early and late type galaxies. Since we are only using late types 
with $b/a>0.6$, we should infer the total number of late types given 
the number of those with $b/a>0.6$. Some late types with $b/a<0.6$ must have 
dropped out of the volume-limited sample due to the internal absorption. 
In Paper I we showed that the correction factor can be reasonably made 
using the factor found in the faintest volume-limited sample of galaxies 
with $M_r < -17.5$.
The error bars are estimated from 16 random subsets of galaxies.
%(***** plot the results without spiral galaxies with b/a<0.6 & compare.. 
%We may have to throw away all spirals with b/a<0.6 <--- check how much 
%is M affected by inclination)

We find that the early type fraction is a monotonically increasing
function of local density at fixed luminosity. It is surprising to
find that this monotonic dependence holds over 3 orders of
magnitude in local density, from the highest down to the lowest
density we explore, and in every luminosity bin. 
The early type fraction falls well below 20\% at extreme low densities.
This late type dominance at low density is particularly true for fainter galaxies.
To test for the
effects of the finite bin size in ${\rm log} \rho$ and $M_r$ on the
relation we also examine the relation using smaller bin sizes.  The
thin solid line between the lines of $M_r = -19.9$ and $-20.4$ is the
morphology-density relation of galaxies in the $D5$ sample with a 
median magnitude of $M_r=-20.25$ using the bin sizes $\Delta
\rm{log} \rho = 0.6$ and $\Delta M_r = 0.5$.  The amplitude and slope
of the relation is quite consistent with those for which larger bin
sizes are used; thus, our results are not affected by the finite bin
sizes.
We also examine the effects of including those galaxies whose
redshifts are borrowed from the nearest galaxy with redshift within 
$55\arcsec$, and find that the dependences of morphology fraction on density and
luminosity are essentially the same.

Our results on the morphology-density relation are consistent with,
and extend to a larger range of density and luminosity,
the findings of several earlier investigators.
For example, in studies limited to rich cluster environments,
Oemler (1974) and Dressler (1980) found higher fractions of E
and S0 types (both classified as ``early type'' here) in the centers
of clusters. Subsequent investigations have verified and refined the
understanding of this effect. As originally noted by
Postman \& Geller (1984) and by a number of more recent investigators
(Lewis et al. 2002; Goto et al. 2003;
Balogh et al. 2004a; Weinmann et al. 2006;
Quintero et al. 2007) this
segregation is not limited to rich clusters, but occurs in poorer
groups as well.  Figure 5 reflects this fact in the strong dependence
of the early type fraction on density even at the lowest densities.  The
dependence of the early type fraction on absolute magnitude, even at fixed
density, has been previously noted (Balogh et al. 2004b;
Kauffmann et al. 2004; Hogg et al. 2004).
As we will discuss later in more detail, the
dependence of the late type subclass on density has also been seen as a
slight dependence of the median blue sequence color on environment
(Balogh et al. 2004b; Blanton et al. 2007). The main difference between our
results here and these previous studies is that most of the previous
studies use either visual classification or use simple classifications
according to optical color.  Although our classification system is
related to broadband optical color, using the color gradients and
concentrations yields a cleaner sample morphologically, bridging the
gap between the studies based on visual classification and those based
on purely color classification.
Another important difference is the use of  spline kernel to define
local density. The cylindrical kernel, which is frequently used
in past studies (Goto et al. 2003), is very noisy and cannot distinguish among 
low density regions well while the local density estimated by the 
 spline kernel holds a good tight correlation with the true density 
field down to extreme void environments (C. Park et al. 2007, in preparation).

Figure 6 presents the 
morphology-density-luminosity relation in a different way. 
Here we find that the 
early type fraction is a monotonically increasing function of 
luminosity at a given local density.
In slightly over-dense regions, where the statistics are very good,
it can be clearly seen that the early type fraction rises abruptly
at magnitudes brighter than $-21.2$ (see also Fig. 9 of Paper I).
The slope of the $L$-$\sigma$ relation also has a break at a similar
magnitude (see Fig. 14, below).
This may imply that many early type galaxies more than about 1 mag
brighter than $M_\ast$ have origins that are different from fainter ones.

It is remarkable that the early type fraction$-$absolute magnitude
relation has similar slope for galaxies with local density ranging
over nearly 3 orders of magnitude. 
At intermediate magnitudes, near $M_\ast$, the early type fraction is
well described by the morphology-density-luminosity formula
\begin{equation}
f(E)=[0.33-0.074(M_r + 20.3)](\rho/\bar{\rho})^{0.164}
\end{equation}
over the whole density range explored.
At magnitudes fainter than about $M_\ast +1$ the local
density dependence monotonically steepens at high density, but becomes 
weaker in under-dense regions (see two bottom curves of Fig. 5. 
A result of this is that 
in very high density regions the early type fraction
decreases less steeply at fainter magnitudes than it does at lower densities
(see the top curve of Fig. 6).
Balogh et al. (2004b) measured the fraction of the red population
of galaxies within fixed absolute magnitude bins as a function of
projected local density $\Sigma_5$ adopted by Balogh et al.(2004a).
Even though galaxies are divided by a different criterion
and the local density measure is different, the local density
dependence they found is qualitatively quite similar.
We emphasize that it is important to exclude highly inclined galaxies
in this analysis. Otherwise, the number of late type galaxies brighter than 
$M_\ast$ is significantly underestimated, and the slope of the early type 
fraction versus absolute magnitude relation is systematically biased high
at bright magnitudes.
\begin{figure}
%\epsscale{0.8}
\plotone{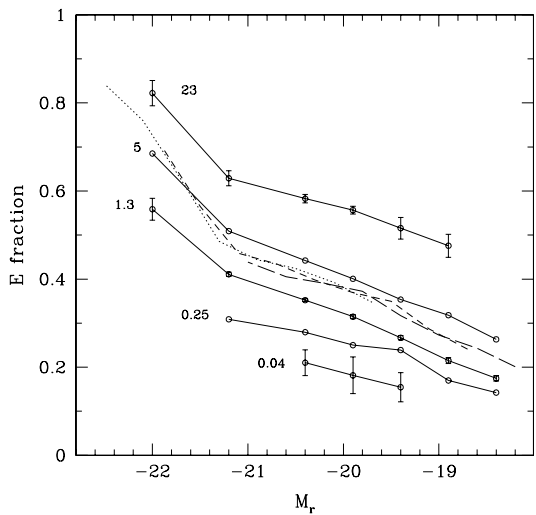}
\caption{Early type fraction as a function of absolute magnitude. 
Labels on solid curves indicate local density, as in Fig. 5.
The long-dashed, short-dashed,
and dotted lines are the morphology-luminosity relations obtained from
the volume-limited samples D1, D2, and D3, respectively, averaged
over all environments.
}
\end{figure}

Next we look for dependence of subclasses of late type galaxies on 
environment. Subclass classification of late type galaxies can be made 
in the color-color gradient space. We divide the late type galaxies into 
two subclasses; late late types with $u-r<1.8$ and early late types with 
$u-r\ge 1.8$. This divides late type galaxies roughly into those of 
Sc type and later and those earlier than Sc, respectively (Park \& Choi 2005). 
\begin{figure}
%\epsscale{0.8}
\plotone{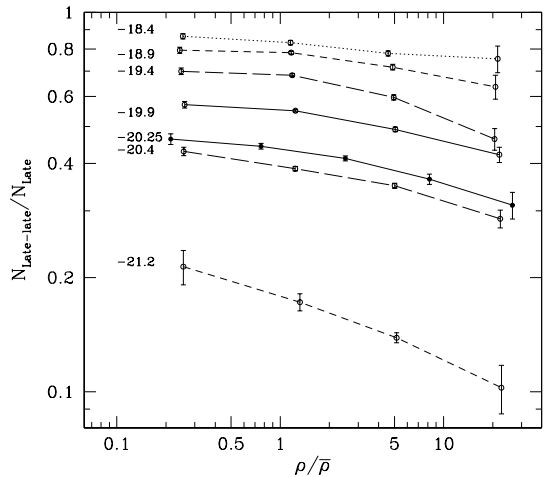}
% MSV: Need to make clear that this is fraction of spirals that are
% late-late, not the overall fraction that are late-late. Figure caption
% says this, but people read too quickly.
\caption{Fraction of late types (of late types, not of all galaxies) 
that are classified as
late spiral type galaxies ($u-r$ color 
bluer than 1.8), as a function of local density at fixed absolute magnitude
(the number attached to each line). The bin sizes are $\Delta M_r = 0.8$
and $\Delta {\rm log} (\rho/\bar{\rho}) = 0.8$,
and points are at the median density and the median magnitude of galaxies
belonging to each bin.
}
\end{figure}
The bin sizes are again $\Delta M_r=0.8$ and 
$\Delta {\rm log} (\rho/\bar{\rho}) = 0.8$, and only the late types
with axis ratio greater than $0.6$ are used.
In Figure 7 we find that the fraction of late spiral types
with $u-r$ color bluer than $1.8$ increases strongly as the luminosity
decreases.
This can be easily understood from the color-luminosity relation of
late type galaxies (see Fig. 3$a$ of Paper I).
This figure also shows that the late spiral population increases
in low density regions for bright spirals, in particular.
Even though our sub-spiral class division is very crude,
it can be seen that the sub-type of spiral galaxies does depend
on local density at fixed luminosity.

\subsection{Luminosity}

%The $r-$ band absolute magnitude $M_r$ is the AB magnitudes 
%converted from SDSS magnitudes. 
Figure 8 shows the dependence of luminosity 
of galaxies in the D4 sample on the local density. The upper panel is for the 
early types and the lower is for the late types with $b/a > 0.6$.
It is clear that high density regions contain very bright galaxies 
while low density regions do not. This indicates that there are systematic 
effects of local density on the maximum luminosity that can
be expressed by a change of luminosity function.
This environmental dependence of the luminosity function was
found by Park et al. (1994) in the CfA redshift survey data, and is now 
confirmed by the SDSS data over 4 orders of magnitude of local 
density. Also evident in Figure 8 is that this local density dependence 
of luminosity is stronger for the early type galaxies. 
The dependence is weaker but can be still observed for the late types.

The distribution of galaxies in the vertical direction at a given
local density in Figure 8 reflects the environment-dependent
luminosity function
\begin{figure}
\plotone{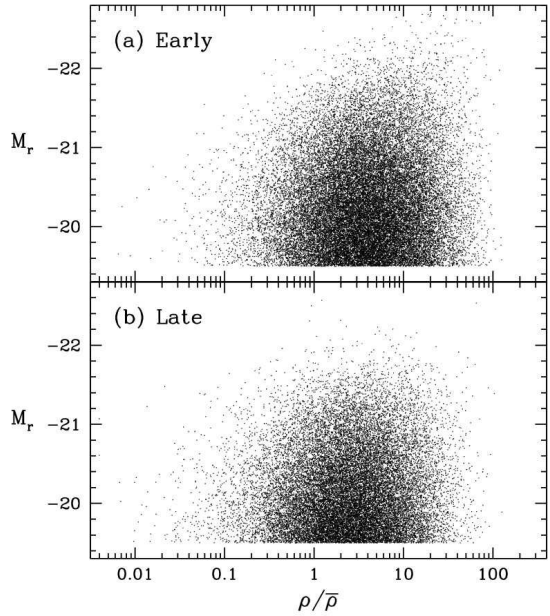}
\caption{Local density dependence of luminosity of early and late
type galaxies in the D4 sample. Very bright galaxies inhabit only high
density regions while faint ones populate all densities. The result is the local
density dependence of the luminosity function.
}
\end{figure}
which we plot in Figure 9$a$. 
% MSV: Add label to figure legend to indicate that curves are for
% different bins of \log\rho/\bar{\rho}
\begin{figure}
\epsscale{1.}
\plotone{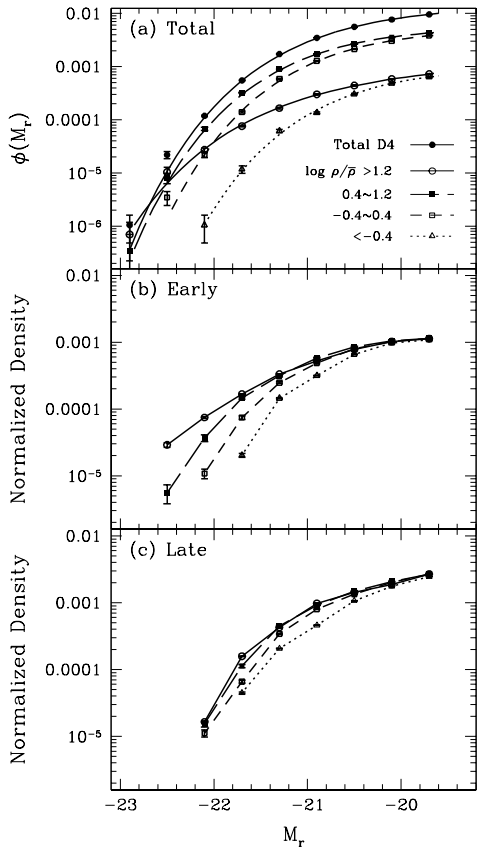}
\caption{($a$) Luminosity functions of all galaxies in the D4 sample
({\it upper filled circles}) and of those located at different local density.
The ranges of ${\rm log} \rho/{\bar \rho}$ are given in
the legend at the lower right corner.
($b$) Luminosity functions of early type galaxies at different 
local densities normalized to a value at the faintest magnitude.
Points and lines have the same meaning as in ($a$).
($c$) Same as ($b$) but for late type galaxies
}
\end{figure}
The top points and curve are the luminosity function measured from the
D4 sample ($M_r < -19.5$) and its best-fit Schechter function,
respectively.  Curves lying below are the luminosity functions of
galaxies located in bins of different local density, normalized to the
whole volume of D4, so that the sum of these curves gives the total
luminosity function.  We exclude late type galaxies with
$b/a<0.6$ in the LF measurements to reduce the bias due to the
internal absorption, and we weight the remaining late types by the
inverse of their fraction.  The LF at the highest density ({\it lower solid
line with open circles}) extends to magnitudes much brighter than those
reached by the LF in the lowest density environment ({\it dotted line with
triangles}), in agreement our qualitative conclusion from Figure 8.  To
answer the question of whether or not the dependence of the LF of
galaxies on local density is due simply to the dependence of
morphology fraction on local density or to a change of the shape
of the LF, we normalize the LFs at the faint end and compare their
shapes.  In each density bin used in Figure 9$a$ we divide galaxies into
early and late type groups, and recalculate their LFs.  Figure 9$b$ and
9$c$ show that the shape of the LF of each morphological type does
change as a function of local density.  In particular, the LF of early
types is more sensitive to environment than that of late types.
Therefore, the local density dependence of the LF is due both to
the change of the early type fraction and that of the shape of the
LF of each morphological type.

The visual results of Figure 9 can be quantified by comparing the
best-fit Schechter function parameters. We divide the galaxies in
the D3 sample ($M_r < -19.0$)
into late and early types and into seven different local
density bins. The solid lines with filled circles in Figure 10 are the
Schechter function parameters $M_\ast$ ({\it upper panel}) and 
$\alpha$ ({\it lower panel}) when the morphology subsets are combined.
\begin{figure}
\epsscale{1.}
\plotone{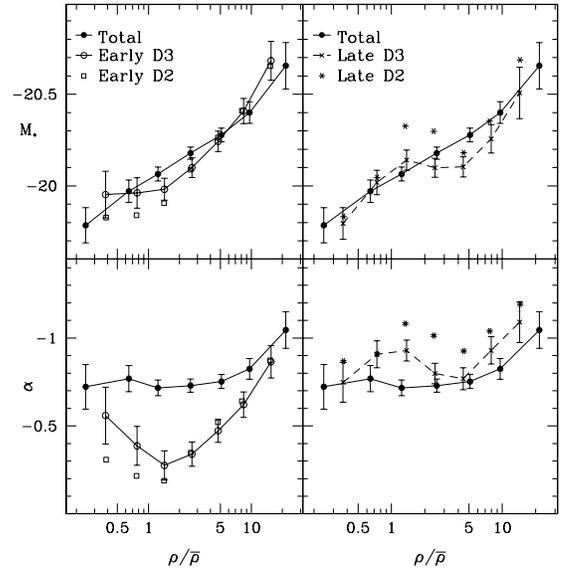}
\caption{Local density dependences of the Schechter function
parameters $M_\ast$ and $\alpha$ calculated from the D3 sample
Filled circles are the results from the whole D3 sample (with 
the inclined late types excluded), and open circles and crosses are
those from the early and late type subsets, respectively.
The symbols ({\it squares and stars}) 
without connecting lines are those from the D2 sample ($M_r\le -18.5$).
}
\end{figure}
We use the MINUIT package of the CERN program library\footnote{
\url{http://wwwasdoc.web.cern.ch/wwwasdoc/minuit/minmain.html}}  
to estimate the
parameters by the maximum likelihood method.  
The estimated Schechter function parameters are listed in Table 2.
The solid line with open
circles and the dashed line with crosses in Figure 10 
are those of early and late
types, respectively. The luminosity of early type galaxies 
as measured by the characteristic absolute magnitude, $M_\ast$, 
starts to rise steeply for local density above the
mean density while that of late types rises below the mean density,
stays more or less constant at intermediated density, 
and then rises again in high density regions.  
Previous work has also found that the high-luminosity turnover in the
luminosity function strongly depends on local density or cluster
richness (e.g., Lumsden et al. 1997; Driver et al. 1998;
Marinoni et al. 2002; De Propris et al. 2003; Balogh et al. 2004a;
Hoyle et al. 2005; Popesso et al. 2005; Zandivarez et al. 2006).
Croton et al. (2005) measured the Schechter function parameters
for early and late type galaxies in the 2dFGRS. 
Contrary to Croton et al., 
we find that $M_\ast$ for early types is changing slowly below
the mean density and that luminosity for late types rises at 
high density ($\rho/{\bar \rho}>5$). The difference seems to be due
to difference in morphology classification. They used
a morphological classifier based on fiber spectra of galaxies.
Morphology classifiers based simply on total
color or spectrum tend to misclassify red spirals into early
type, and blue ellipticals into late type.
%Decrease of $\alpha$ parameter of early-types' LF in low density
%regions is caused by the blue ellipticals, and will not be
%noticed unless the blue ellipticals are correctly classified.
The color gradient and concentration constraints in our
morphological classification strongly reduce such misclassifications.
Another important difference in our analysis is the exclusion of
inclined ($b/a<0.6$) late type galaxies. When they are included, the
number of relatively fainter galaxies is overestimated because many 
of the inclined galaxies are actually bright, but appear fainter due
to internal absorption. In particular, this systematic effect of
internal absorption makes the $\alpha$ parameter of late types
more negative and makes the $M_\ast$ of late types appear fainter.  
The rise of the characteristic luminosity of late types 
in high-density regions would not be detected unless the bright 
red spirals, which are abundant at high density, are correctly 
classified and the internal absorption effects are eliminated.

The faint-end slope, $\alpha$, does not show a monotonic behavior
for late type galaxies, consistent with the results of 
Hoyle et al. (2005) and Croton et al. (2005).  
%At high density, however, $\alpha$
%starts to decrease for both early and late type galaxies.
%Hoyle et al. and 
%Rojas et al. (2004) have shown that their void galaxies in the SDSS are 
%also typically bluer than those in the high density environment. 
% MSV: Changbom - please check. I replaced the following sentence
% with two new sentences below. Please also check my changes to discussion
% in this paragraph
%
%It is interesting to see that $\alpha$ of the early type's LF first 
%increases as the local density decreases and then decreases at 
%local density below the mean (see bottom curve of the bottom panel 
%of Figure 10). 
It is interesting that $\alpha$ for early types decreases (steeper
faint end slope) at local density both below and above the mean.
At local density near the cosmic mean, the
early type LF has an exceptionally flat faint end slope: $\alpha\sim
-0.3$.  Detection of this behavior requires careful morphological
classification, because faint early type galaxies reside
predominantly not only in massive halos (which prefer high density
environment) but also in low mass halos present in under-dense regions.
The latter tend to be fainter bluer early type galaxies that are
correctly classified by our morphological classifier. If galaxies
were classified by color or their spectral properties alone, these
blue early types would have been classified as late types.  
We emphasize again that the major differences between our analysis and
the previous ones lie in the morphological classification and
exclusion of highly inclined late type galaxies, as well as the new
local density estimator.
On the other hand, one should be cautious about the selection bias when
the abundance of different species of galaxies are compared. 
In the case of the SDSS catalog 
there might be some bias due to the fact that  
some very low surface brightness galaxies are missed.

The Schechter function parameters are measured from our 
volume-limited samples with magnitude limits of $-19.0\pm 1.0$.
In Figure 9 we used the D4 sample ($M_r \ge -19.5$) to better see 
the behavior of the bright tail of the LF, and in Figure 10 the D3 sample 
($M_r \ge -19.0$) was adopted because this sample has the minimum 
uncertainties when both $M_\ast$ and $\alpha$ are concerned.
However, these samples whose ranges in absolute magnitude are only
about 1 mag below $M_\ast$, are not really proper for giving accurate
estimates of $\alpha$. The limited range in absolute magnitude also
develops some correlation between the estimates of $M_\ast$ and $\alpha$.
This is particularly so at low density, where $M_\ast$ is fainter,
and for late types for which the faint end slope of the LF is steeper.
We compare the local density dependences of LF parameters
measured from all five volume-limited samples with one another 
to see the effects of finite magnitude range.
In Figure 10 the results from the D2 sample ($M_r\le -18.5$) are plotted
as symbols without connecting lines for a comparison.
We confirm that the behaviors of $M_\ast$ of early and late type 
galaxies as a function of local density were basically the same
for all samples. It was also true for the $\alpha$ parameter
of early types. However, the dependence of $\alpha$ of late types
on local density showed the systematic trend that $\alpha$ became
more negative as the faint limit of the sample became fainter.
In order to measure $\alpha$ of late types accurately, one needs
to use fainter galaxies, and for a flux-limited sample this makes 
one rely exclusively on the very close faint galaxies.
It should be noted that the nearby universe is a relatively 
under-dense part of the universe, and that the resulting parameters
can be also biased due to the local density dependence of the LF.
Dependence of $\alpha$ on environments is
still a matter of some debate
(e.g., de Propris et al. 1995; Secker et al. 1997; Valotto et al. 1997, 2001;
Trentham 1998; Andreon \& Cuillandre 2002; Popesso et al. 2006).

\subsection{Color}

It is well known that galaxy colors are redder in regions of higher
local density
% (e.g. Butcher \& Oemler 1984; Smail et al. 1998;
%Ellingson et al. 2001; Margoniner et al. 2001; Zehavi et al. 2002;
%Balogh et al. 2004a; De Propris et al. 2004; 
(Hogg et al. 2004; Blanton et al. 2003c; Tanaka et al. 2004).
In this section we show this trend is
dominated by the dependence of color on luminosity and morphology. 
For this purpose it was necessary 
to examine the dependence of color on local density
when both morphology and luminosity are fixed.
We use $u-r$ color in this work because it is a measure of the star
formation activity of galaxies in the recent past.  
The top panels of
Figure 11 show scatter plots of $u-r$ color of early ({\it left panels}) and
late ({\it right panels}) type galaxies in the D2 sample
whose absolute magnitudes are between $M_r=-18.5$ and $-19.3$
(the median magnitude is $-18.9$).
%That is, they show the color versus local density relation at
%a fixed luminosity. We also study the relation for fixed morphology by
%showing earlytypes in the left panel, and late types in the right panel. 
The thick blue solid lines delineate the most probable
(mode of the distribution of color at fixed absolute magnitude and
at fixed local density interval) 
color of early types and the median color of late types. Late type galaxies
with axis ratios $b/a <0.6$ are again excluded.
%Other lines are for galaxies in brighter luminosity bins. For example,
The short-dashed green lines are the most probable (early) and
median (late) colors of galaxies with $-19.3>M_r>-20.1$.
The corresponding scatter plots are not shown to avoid confusion.
The long-dashed magenta line and the red solid line are for galaxies
in magnitude bins of $-20.1>M_r>-20.9$ and $-20.9>M_r>-21.7$
with median magnitudes of $-20.4$, and $-21.1$, respectively.

\begin{figure*}
%\hbox{\hspace{2.0in}Early \hspace{2.4in} Late}
%\epsscale{1.2}
\plotone{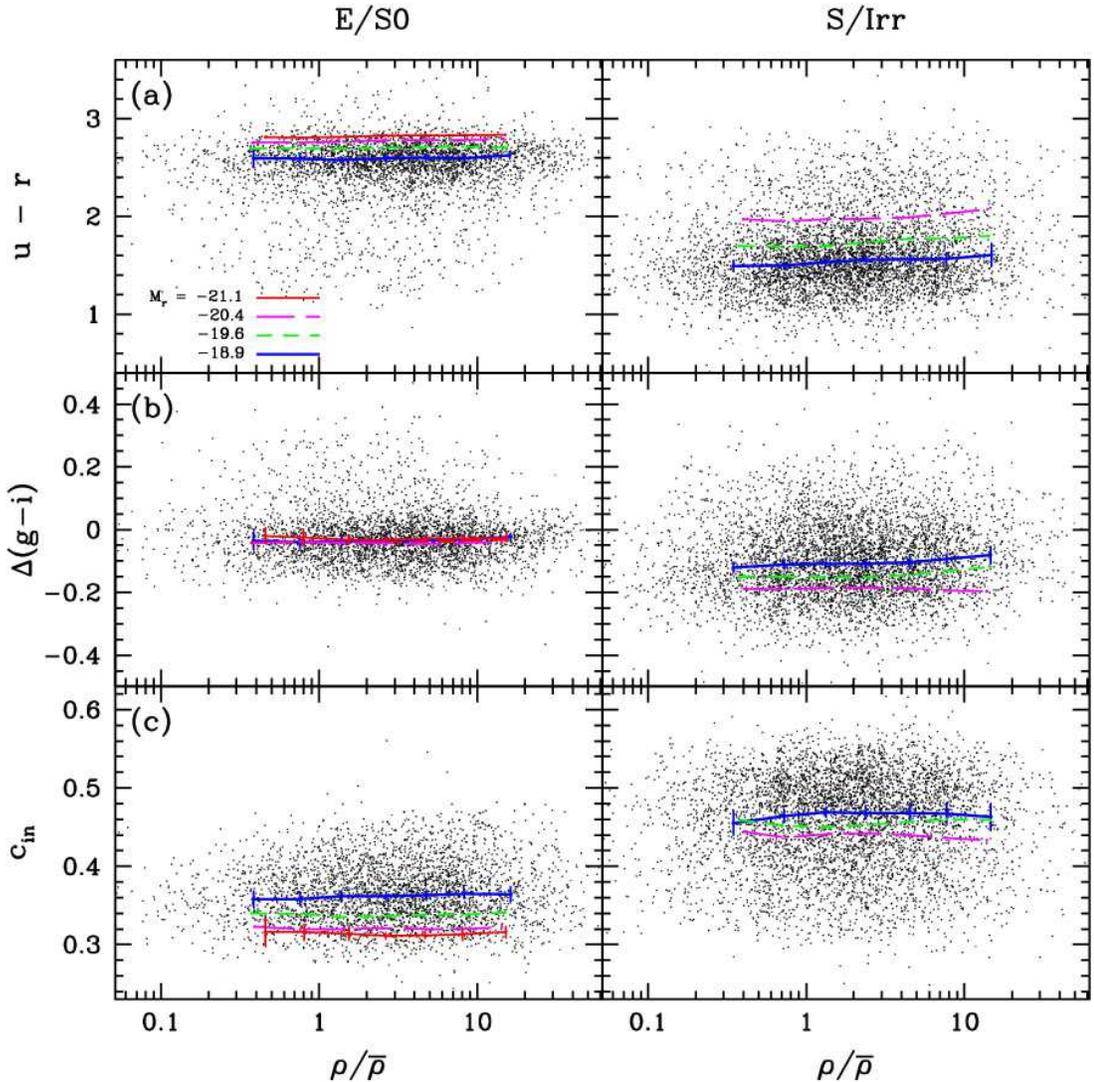}
\caption{
Local density dependence of ($a$) $u-r$ color, ($b$) $\Delta(g-i)$ color
difference, and ($c$) $c_{\rm in}$ (inverse) concentration index.  Dots
are galaxies in the D2 sample with $-18.5 >M_r > -19.3$.  The early
type galaxies are on the left, and late types are on the right. Blue
solid lines are the median curves except for the early types in panels
($a$) and ($b$) where the most probable values are drawn. Other colored lines
show local density dependences at brighter magnitudes of $M_{r}
\approx -19.6, -20.4$, and $-21.1$. Uncertainty limits are attached
to the relations for the faintest and the brightest subsamples.}
\end{figure*}

We first note that there is no break in color as a function of local
density for either early or late type galaxies. Tanaka et al. (2004)
reported detection of a break in color at a density of $\rm{log}\Sigma_5
\sim 0.4$ galaxies $h_{75}^2$Mpc$^{-2}$ (see Tanaka et al. for the
definition of projected local density $\Sigma_5$) for galaxies with
magnitudes $M_\ast +1<M_r<M_\ast +2$. It is clear that this result is simply 
due to the fact that blue late types tend to avoid in very high density
environments, and it is the morphology bias rather than the color bias
that causes the effect. This feature in the morphology bias appears
as a break in the morphology-luminosity relation at an absolute 
magnitude of about $M_\ast-1$ (Paper I).

Second, we note that the color of galaxies is a very insensitive
function of local density once the morphology and luminosity are
fixed.  
As the local density varies from $\rho/{\bar\rho}\approx 15$ to about
0.37, the color changes only by 0.03 for early types and by 0.11 for
late types; both get redder at higher density. This behavior is
insensitive to luminosity,
%(except for the reddest spirals)
which can
be seen from the fact that the lines are nearly parallel to one
another.  On the other hand, the color changes by 0.22 and 0.75 as the
magnitude changes by 2.2 from $M_r=-18.9$ to $-21.1$ for early and
late types, respectively. The difference between the typical colors of
early and late types is as large as 1 mag when $M_r = -18.9$.
Therefore, the dependence of color on local density is dominated by
the dependences of color on morphology and luminosity, which in turn
are strongly dependent on local density, as shown in sections 4.1 and 
4.2. Figure 11 also shows that
blue early types live mainly at intermediate- and low-density
environments, and that some red early type galaxies still exist in the
extreme low density regions with $\rho/{\bar\rho}<0.1$.
% MSV: The next paragraph seems out of place because it is in between
% of Figure 11a and Figures 11b, c. Perhaps Figure 12 and its discussion could 
% first in the color section, before the current Figure 11?
Tanaka et al. (2004) reported that the star formation rate indicated 
by the $g-i$ color and
the equivalent width of the $H\alpha$ emission line is suppressed in
dense environments.
It turns out that most of the effects they detected are due to the
dependence of morphology and luminosity on local density.

To inspect the collective behavior of color of each morphological
type we examine the early and late type sequences 
in the color-magnitude diagram. The points in Figure 12 
are early ({\it left panels}) and late ({\it right panels}) type 
galaxies in the D3 sample
located in high ({\it upper panels}) and low ({\it bottom panels}) 
density environments. 
\begin{figure}
\epsscale{1.}
\plotone{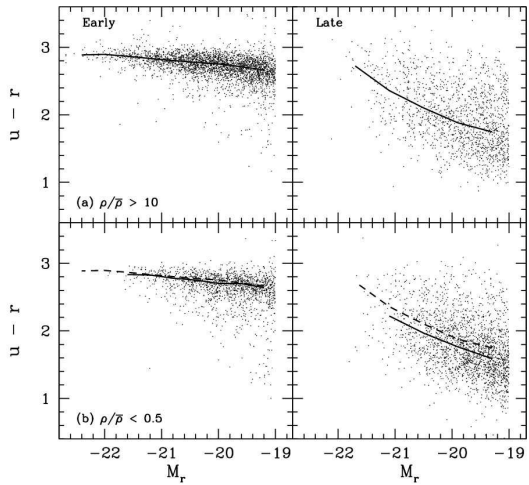}
\caption{Dependence of the red and blue sequences in the color-magnitude
diagram on the local density. Solid curves are the most probable (early)
and the median (late) relations defining the sequences at different 
environments. The solid curves in the upper panels are plotted again
in the lower panels as dashed curves for comparison.
}
\end{figure}
In the upper panels, solid lines are the most probable color for the
early types and the median color for the late types.  In the bottom
panels, the most probable and median color curves at high density are
replotted as dashed lines, for comparison with the curves at lower
density. We find that the shift of the early type sequence in the
color-magnitude diagram toward bluer color is only 0.025 mag
when the local density changes from $\rho/{\bar\rho}>10$ to $<0.5$.  A
larger shift of 0.14 in $u-r$ color is observed for the late type
sequence.

The very slight dependence of the location of the red sequence on the
density field has been noted in the past, originally by
Sandage \& Visvanathan (1978), but more recently by others
(Balogh et al. 2004b; Hogg et al. 2004; Bernardi et al. 2006; 
Blanton et al. 2007). The stronger dependence of the
blue sequence colors on environment was pointed out by
Balogh et al. (2004a). 
%Our results confirm that this effect is not likely to be due to 
%%an increase in dust content (since we have restricted to
%%galaxies with $b/a>0.6$) or other 
%effects of misclassification in dense regions.

\subsection{Color Gradient}

The points in Figure 11$b$ are the color gradient of galaxies in
the D2 sample as a function of local density.  The most probable color
gradient of early type galaxies is slightly negative (the center is
redder) and shows no environmental dependence at a given absolute
magnitude. It is nearly independent of luminosity as well. This
extraordinary result is due to the fact that the internal color
gradient of early type galaxies is almost independent of luminosity
even though the color itself does have a monotonic dependence on
luminosity (cf. Fig. 3$a$ and 3$b$ of Paper I).  Late type galaxies have
more negative color gradients when they are brighter (cf. Fig. 3$b$ of
Paper I), which explains the dependence of the color gradient of late
types on absolute magnitude in the right panel of Figure 11b.  
This may reflect a change in the bulge-to-disk ratio with luminosity.
Figure 11$a$
shows that fainter late type galaxies are bluer in their integrated
color than bright ones, and Figure 11$b$ indicates that they have bluer
centers (thus lower color gradients). The color difference weakly
depends on local density for faint late type galaxies while no
dependence is seen for bright galaxies.
%(also reflected in the
%differences in concentration we find in the next subsection).  
%This
%aspect of the galaxy's morphologies appears not to be related to the
%environment.

\subsection{Concentration Index}

At fixed color or estimated star-formation history, several previous
authors found little dependence of concentration on environment
(Hashimoto \& Oemler 1999; Bernardi et al. 2003a; Kauffmann et al. 2004;
Blanton et al. 2005a; Quintero et al. 2007).
We extend these results by showing that even when
we use a cleaner definition of morphology, the radial profiles of
galaxies appear to be only weakly related to their environment.

When morphological type and luminosity are fixed,
the surface brightness profile, represented by the (inverse)
concentration index, is nearly independent of local density as shown
in Figure 11$c$.  The median $c_{\rm in}$ of bright early types with
absolute magnitudes within 0.8 mag bins shows little change at
all local densities explored. Therefore, the dependence of surface 
brightness profile of early types on local density is solely due to 
the dependence of the profile on luminosity (cf. Fig. 3c of Paper I), 
which monotonically depends on local density, as demonstrated in 
section 4.2.  Faint early types seem to be slightly
more concentrated at low density.
It is difficult to see a dependence of concentration index of late
types on local density because the dispersion in $c_{\rm in}$ is
large and the effects seem weak.

\subsection{Size}

% MSV: Hard to see these results in a log R plot - does Figure 13a look
% better when linear R is plotted?

The size (Petrosian radius) of galaxies is an increasing function 
of luminosity as shown in Figure 4$a$ of Paper I. 
Figure 13$a$ shows that physical size is almost 
independent of local density when luminosity and morphology 
are fixed. We detect a small monotonic dependence of galaxy
size on local density for both early and late types, except for the very
bright galaxies. Surprisingly, galaxies of the same luminosity
are smaller and therefore have higher surface brightness at low density.
Galaxies with $M_r = -19.7\pm 1.2$ located at low density 
($\rho/{\bar{\rho}} \approx 0.4$) are about 8\% smaller in size 
than those at high density ($\rho/{\bar{\rho}} \approx 15$), 
for both early and late types.
The weak dependence of size on local density should be accepted 
with caution, because
including light from neighboring galaxies can in
principle create a trend in the sizes with density
in the direction we see. 

Kauffmann et al. (2004) and Blanton et al. (2005a) have both investigated the
relationship between galaxy size and environment (at fixed luminosity
and at fixed star-formation history or color, respectively), and both
found only small effects. In the range of luminosities
we consider here, more luminous than $M_\ast$, Blanton et al. (2005a) find
a trend for larger galaxies to be in denser regions as well (although
they found this trend is reversed at lower
luminosities). Bernardi et al. (2003b) investigated the dependence
of residuals of early type galaxies from the fundamental plane on
environment and find that the residuals from the fundamental plane in
effective surface brightness correlate slightly with environment. The
sense is that at fixed luminosity and velocity dispersion, galaxies are
larger in dense regions, similar to our results here.

\begin{figure*}
%\epsscale{1.05}
%\hbox{\hspace{2.0in}Early \hspace{2.4in} Late}
\plotone{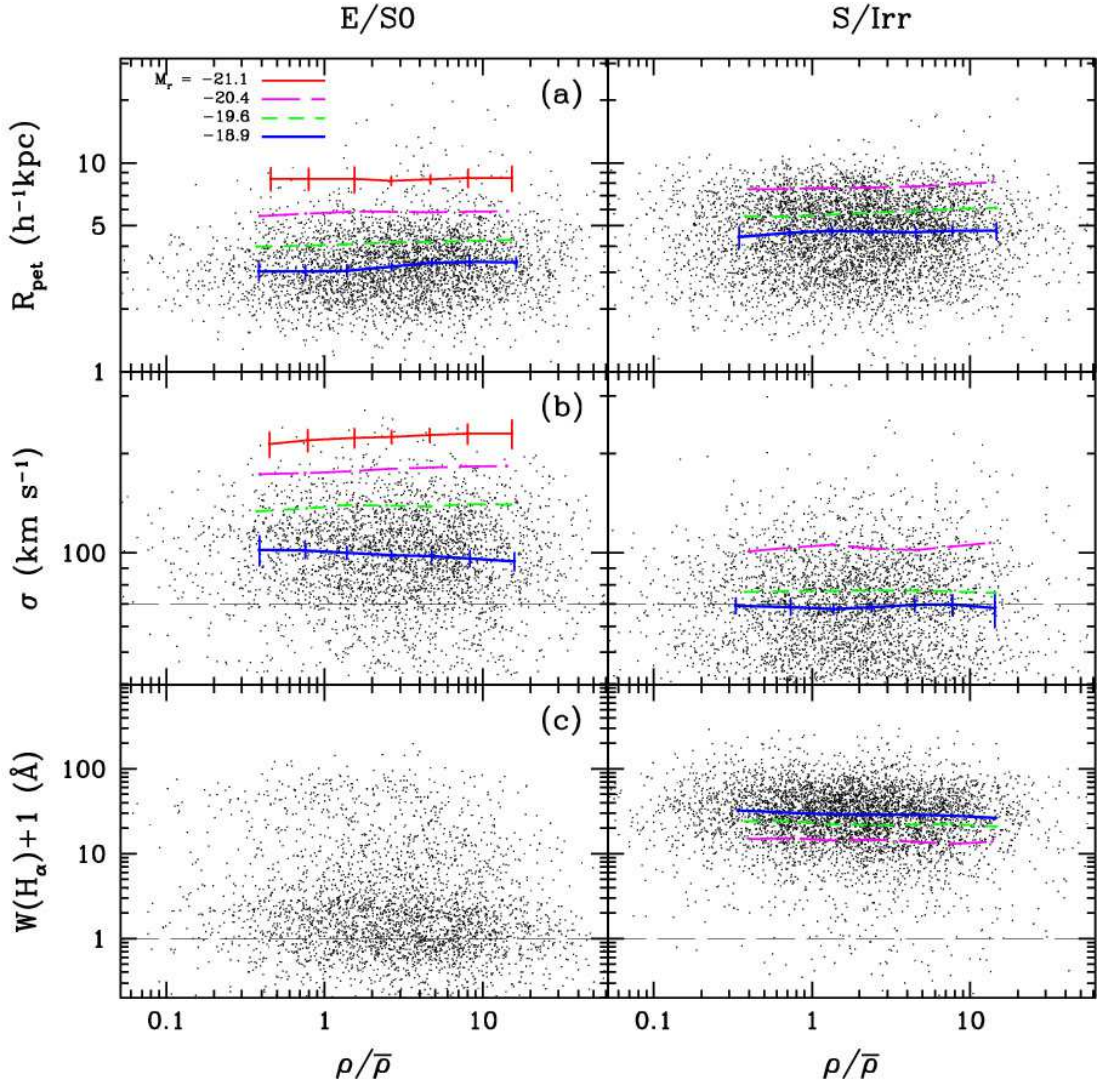}
\caption{Same as Fig. 11, but for ($a$) the seeing-corrected elliptical
Petrosian radius, ($b$) the central velocity dispersion, $\sigma$, and ($c$) the
equivalent width of the $H\alpha$ line. All lines are the
median relations.
}
\end{figure*}

\subsection{Velocity Dispersion of Early Type Galaxies}

In the right panel of Figure 13$b$ we
see that the central velocity dispersion of late type galaxies shows
no dependence on local density.  However, we find an interesting
dependence of the $L$-$\sigma$ relation on environment for early type
galaxies.  The velocity
dispersion of fainter early types slightly increases toward low-density
regions, while the opposite is true for brighter early types.  This might
be evidence for an environment-dependent Faber-Jackson relation or,
more generally, such a dependence of the $L$-$\sigma$ relation.

To explore this possibility further we plot early type galaxies in
high and low-density regions in the absolute magnitude versus velocity
dispersion space. 
The left panels of Figure 14 show the scatter plots of early type
galaxies in the D4 sample located at different density and their most probable
relations in the velocity dispersion-magnitude diagram.  The dashed
curves in the lower panel are equal to the solid ones in the upper
panel. We clearly see that relatively bright early type galaxies in 
higher density regions have higher central velocity dispersions than
those in under-dense regions.  The right panels are for the galaxies in
the D2 sample and show the relations down to fainter magnitudes.
These plots indicate that the slope of the $L$-$\sigma$ relation
depends on local density and that this trend is more evident at faint
magnitudes.  If we fit a power-law relation $L \propto
\sigma^{\gamma}$ to the curves of most probable velocity dispersion,
we measure $\gamma = 1.8$ and $2.2$ in regions with local densities of
$>10$ and $<0.5$, respectively, near $M_r = -19.4$.  Thus, not only is
the relation between luminosity and $\sigma$ not well represented by a
single power-law (cf. Fig. 4$b$ of Paper I), the best-fitting such
relation depends on local density of the galaxies.

\begin{figure}
\epsscale{1.}
\plotone{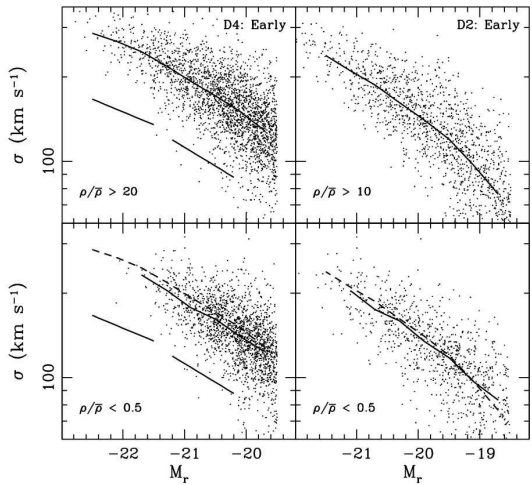}
\caption{Central velocity dispersion vs. absolute magnitude
relations at high ({\it top panels}) and low ({\it bottom panels}) density
environments.
Solid curves are the most probable relations 
(the dashed curves in bottom panels are the same as the solid curves 
in the top panels).
Straight bars illustrate the slopes of the $L$-$\sigma$
relation of $\gamma=4$ ({\it left}) and $\gamma=3$ ({\it lower right}).
}
\end{figure}

\subsection{Star Formation Rate}

The equivalent width of the $H{\alpha}$ line is measured from the
SDSS spectra, and is often used as a measure of recent star formation
activity (or nuclear activity).  
Because SDSS spectra are obtained from the central $1.5''$
radius region of galaxies, this line width is not necessarily a
representative measure of star formation activity, particularly for
spirals with a significant bulge component. Nevertheless, the star
formation activity as measured by the equivalent width of the
$H{\alpha}$ line, is on average stronger for faint late type galaxies
(cf. Fig. 4$c$ of Paper I).
%Its local density 
%dependence is negligible even though *****? One thing to note for early 
%types is that star forming early types mainly live in intermediate and 
%low density regions.
We find that $H\alpha$ line strength varies little with local density
once the morphology and luminosity are fixed.
Note that a strong local density dependence of
$W(H\alpha)$ can be obtained if all morphological type
and luminosity are mixed because the fraction of late type
and faint galaxies increases at lower densities.
However, we are able to
detect a weak correlation between $H\alpha$ line width and local
density for late type galaxies; the star formation activity signaled
by the $H\alpha$ line strength is relatively stronger at low density
environment at a fixed luminosity. For galaxies with $M_r = -18.9
\pm 0.4$ a linear fit to the median relation gives
log$W(H{\alpha}) = 1.466-0.046 \log \rho/{\bar{\rho}}$.
This dependence is weaker for brighter galaxies. 
% MSV: I added the following. Add more discussion of this point?
This variation of star formation activity with environment is
consistent with the findings of Rojas et al.(2005), who compared star
formation activity of galaxies in voids with that of galaxies of
similar luminosity in denser environments.

Other investigators who have explored the dependence of H$\alpha$
emission on density have found that even at fixed galaxy concentration
there is a strong relationship between the fraction of emission line
galaxies and density.  However, G\'{o}mez et al. (2003) and Balogh et al.
(2004a) found, as we do here, that the distribution of H$\alpha$ equivalent
widths for emission line galaxies are only weakly related to density.
These results parallel the similarly weak relationship between color
and environment for late type galaxies.
The majority of early type galaxies show vanishing star formation
activity regardless of luminosity. We note that there is a group of
active early types located at median or low density environments.

\subsection{Axis Ratio of Early Type Galaxies}

The points in Figure 15 are the isophotal $b/a$
ratio of early type galaxies in the D4 sample
with absolute magnitudes between $-19.5$ and $-20.3$
as a function of local density. 
Their median relation is the thick solid line.
Other lines are for brighter galaxies in the same sample, and
the corresponding scatter plots are not made, in order to avoid confusion. 
Brighter early types are rounder than fainter ones 
(see also Fig. 5 of Paper I),
which might be explained by the tidal interaction theory 
(cf. Thuan \& Gott 1977).
Once the luminosity is fixed, the local density dependence of the
axis ratio of early types is so weak that it is not detected
in this plot. 
The bottom panel of Figure 15 compares the histograms of the $b/a$
ratio of galaxies at very high and very low density environments.
No statistically significant difference is observed.
\begin{figure}
\epsscale{1.}
\plotone{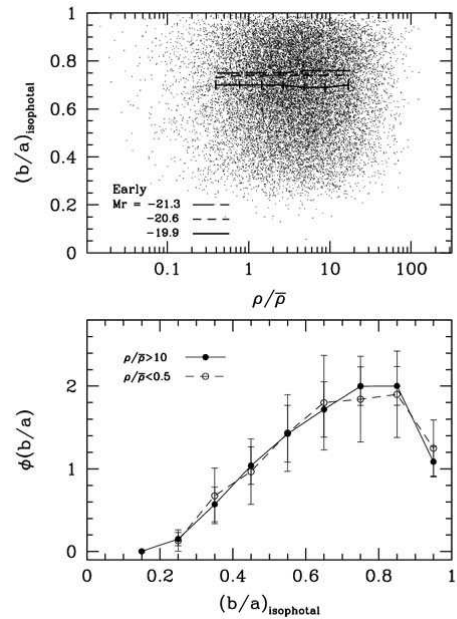}
\caption{{\it Top}: Isophotal axis ratio of early type galaxies 
as a function of local density. 
Galaxies are selected from the volume-limited 
D4 sample ($M_r\leq-19.5$). Dots are galaxies with absolute magnitudes
between $-19.5$ and $-20.3$, and the solid line delineates the median
relation for theses galaxies. The other two lines are for brighter galaxies.
{\it Bottom}: Histogram of the axis ratio of early type galaxies in the
D4 isample located at high and low density environments. 
Only galaxies with absolute magnitudes between $-19.5$ and $-20.3$ are used.
}
\end{figure}

%\subsection{Density, density gradient and density shear}
%
%We have used the local density of $M_*$ galaxies extensively to define 
%environment. However, environment of galaxies can mean more than the local 
%density. As a next step, we adopt the density gradient and shear to 
%distinguish from different environments. We define the density gradient 
%by $\nabla_i = (\partial\rho /\partial x_i ) /\rho$. The density shear is 
%defined by $\nabla_{ij} = (\partial^2 \rho /\partial x_i \partial x_j) /\rho$. 

\section{Dependence on smoothing scale}

So far, we have used local density estimated by the spline kernel
smoothing in which the kernel size varies to include $N_{s}=20$
galaxies in the LS sample of ${\sim}L_\ast$ galaxies.  
We now inspect the environmental
dependence of physical properties of galaxies when the smoothing
kernel includes 200 galaxies.  
Because we have found that most
galaxy properties are nearly independent of local density
once morphology and luminosity are fixed, we look for the
possible dependence of only the morphological fraction and luminosity
function on the much larger scale density field. The isodensity
contours of the upper and lower plots in Figure 4 compare the two
density fields $\rho_{20}/{\bar{\rho}}$ and $\rho_{200}/{\bar{\rho}}$
smoothed over $N_{s}=20$ and $200$, respectively. 

Figure 16 shows the early type galaxy fraction as a function of 
$\rho_{200}/\bar{\rho}$ at each fixed luminosity with an absolute magnitude
bin size of 0.8.  Surprisingly, the nature of the relations 
is basically the same as that found by using a smaller
smoothing scale of $N_{s}=20$ (although the relation seems steeper
for fainter galaxies).  We have checked that we reach the
same conclusion even for $N_{s}=300$.
% MSV: I think the E fraction looks steeper at the faint end when using N_S=200
% Please see the lower curves in Figure 16 and compare to Figure 5.
% This might be the signature of what happens in large-scale voids.
Figure 17 demonstrates the dependence of LF of galaxies on 
$\rho_{200}/\bar{\rho}$. The Schechter function parameters $M_\ast$ and
$\alpha$ depend on the larger scale density $\rho_{200}$ in the same way that 
they do on the smaller scale density $\rho_{20}$(see Fig. 10).  The
characteristic luminosity strongly increases as $\rho_{200}$
rises. The type-specific parameters also behave in the same
way. The only noticeable difference is the $\alpha$ parameter
of the combined sample at the lowest density, but the statistical
significance of that single point is low.
% MSV: Looking at Figure 17, I think the top panel (M\ast) looks the
% same as Figure 10.
% However, the lower panel (alpha) looks different. Here alpha of late types
% is somewhat steeper at high and low density, while alpha of early
% types mostly increases with local density.
It should be noted that the smoothing scale corresponding to
$N_{s}=200$ is not far from the linear scale (the rms density
fluctuations is about $0.69$ while it is $1.8$ for $N_{s}=20$).
The morphology and luminosity of galaxies appear to respond sensitively to
the linear density field.
\begin{figure}
\epsscale{1.}
\plotone{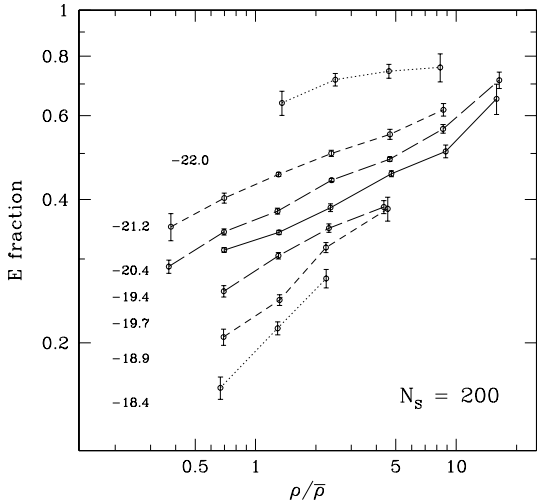}
\caption{Same as Fig. 5, but the spline kernel-weighted background
density is estimated from the 200 nearest $L_\ast$ galaxies at each galaxy
in the D3 sample. 
}
\end{figure}
\begin{figure}
\epsscale{1.}
\plotone{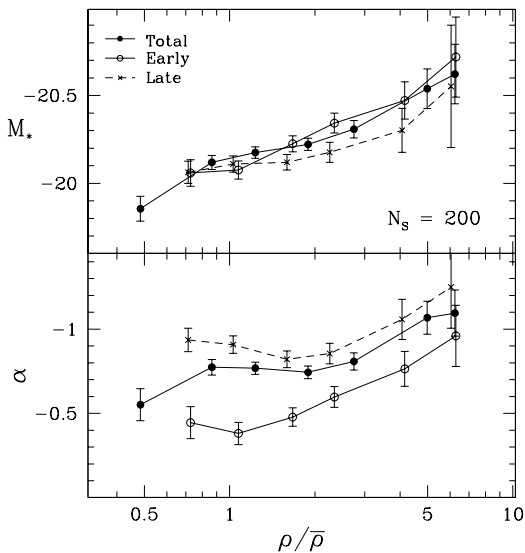}
\caption{Schechter function parameters $M_\ast$ and $\alpha$ as a
function of the large scale background density calculated by using
$N_{s}=200$ in the spline kernel smoothing. Galaxies in the D3 sample are
used.
}
\end{figure}

Since the density field that has $N_s=200$ smoothing is correlated 
with that having $N_s=20$, it is not surprising to find 
similar relations between the local density and galaxy properties
at both smoothing scales.  Therefore, it is now very important 
to identify the scale on which the large-scale environmental dependence
of morphology and luminosity of galaxies originates.
To answer this question we study the dependence of the
morphological fraction on very {\it small} scale environment
by extracting samples both with and without close neighbors.
We select
``isolated'' galaxies brighter than $M_r =-19.5$ in the D2 sample 
D2 ($M_{r,{\rm lim}}=-18.5$). We adopt the isolation
criteria for a target galaxy with $M_r$ that there can be 
no companion galaxy in the absolute magnitude
interval $M_{r,i}+1 >M_r >M_{r,i}-2$, with redshift difference within 300 km s$^{-1}$,
and with separation across the line of sight less than $500 h^{-1}$kpc
for a candidate galaxy $i$.
The solid line in Figure 18 is the early type fraction of isolated
galaxies as a function of $\rho_{20}/\bar{\rho}$.
\begin{figure}
\plotone{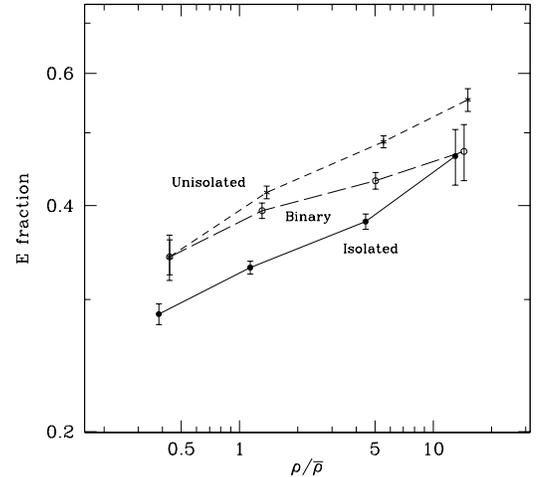}
\caption{Early type fractions of isolated galaxies ({\it solid line})
in the D2 sample brighter than $M_r = -19.5$.
as a function of local density ($N_s = 20$).
The long- and short-dashed lines are
the early type fractions of the galaxies in binary systems or
those with one or more than one companions, respectively.
}
\end{figure}
Since the morphological fraction of isolated galaxies with no close 
companion still correlates tightly with $\rho_{20}$,
we conclude that the large-scale
environment is definitely one of the factors that determine
the morphology of galaxies. 
We reach the same conclusion when the isolation criteria are changed
to 500 km s$^{-1}$ in redshift difference and $1 h^{-1}$ Mpc in the separation
across the line of sight.
In Figure 18
the long and short dashed lines are the early type fractions 
of the galaxies with only one neighbor (binary) or those with 
one or more companions (unisolated), respectively.
%Two lines meet each other at low density because unisolated
%systems in low density regions are dominated by binaries.
We see that galaxies having close companions are more
likely to be of early type than isolated galaxies, even at
the same large-scale density.
This demonstrates that the morphology of galaxies is determined in
response to very small scale density as well as
to large ($N_{s}=20$) and very large ($N_{s}=200$) scale
background density fields.

To narrow down the scale responsible for the origin of morphology
we push our study further using the densities at these three scales
in Figures 19 and 20. 
Figure 19 shows the relation between the
densities estimated with $N_s=20$ and 200 at each galaxy in the D2 sample. 
Superposed are contours of constant early type galaxy fractions
of 0.25, 0.30, 0.35, 0.40, and 0.45 from left to right.
\begin{figure*}
%\epsscale{1.2}
\plotone{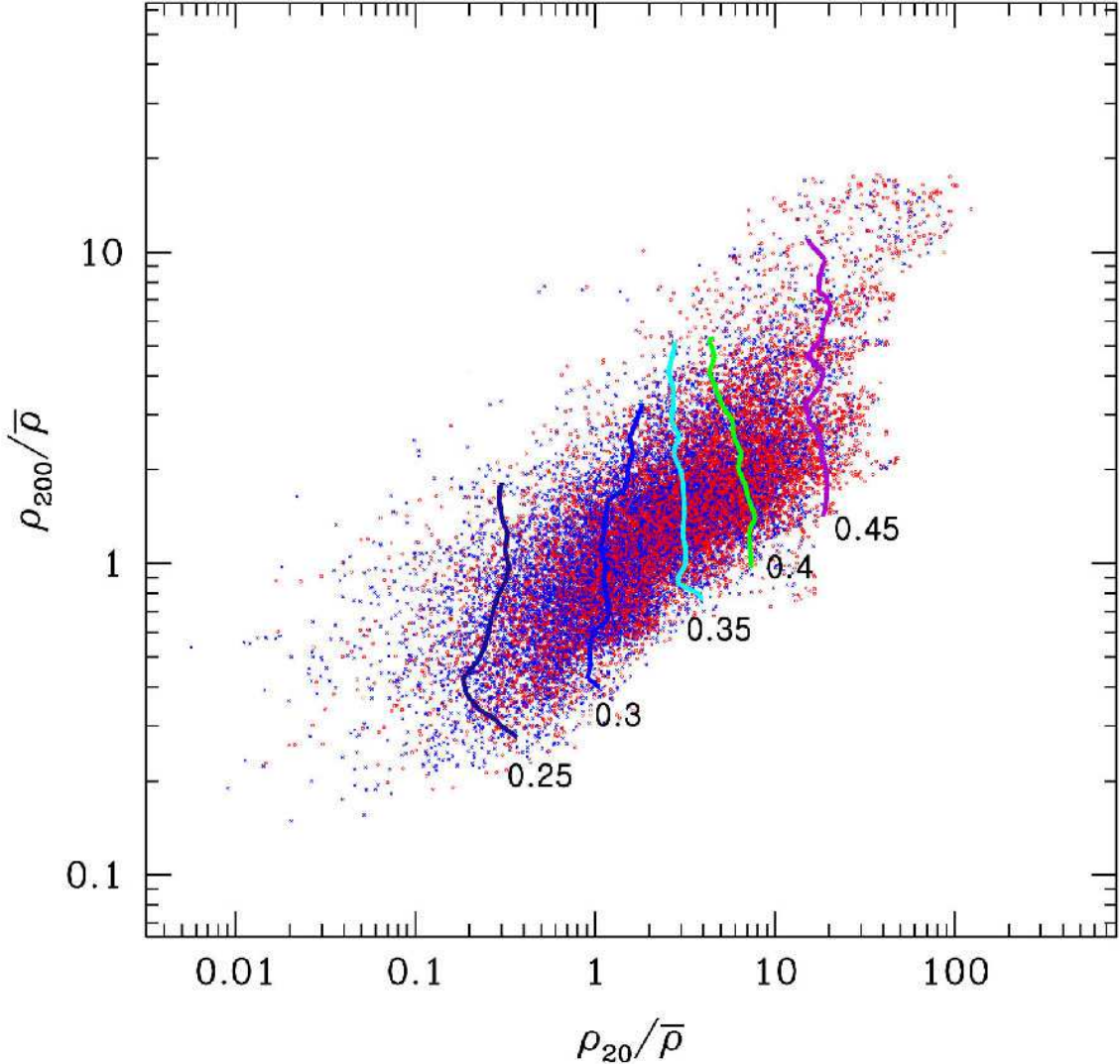}
\caption{Relation between the density $\rho_{20}$ estimated
from $N_{s}=20$ spline smoothing and the density $\rho_{200}$ smoothing.
Galaxies are those with $M_r \leq -18.5$ in the D2 sample. 
Superposed on the scatter plot are the contours of constant early type
fraction of 0.25, 0.30, 0.35, 0.40, and 0.45, from left to right.
}
\end{figure*}
We note that the contours are basically vertical. 
The result shows that the morphology of galaxies sitting at a fixed 
$\rho_{20}/{\bar{\rho}}$ is independent of the larger scale
density $\rho_{200}/{\bar{\rho}}$ and  
that the dependence
of morphology fraction on the density with $N_s=200$ smoothing is merely 
because of its correlation with the density when $N_s=20$.
Therefore, the origin of galaxy morphology
must be due to the environmental conditions at scales much lower 
than the $N_s=200$ scale, which is roughly 
the $12h^{-1}$ Mpc Gaussian smoothing scale.

Figure 20 compares $\rho_{20}/{\bar{\rho}}$ with the distance to 
the nearest neighbor from a galaxy, $r_{\rm near}$, representing
the small-scale density in this logarithmic plot.
\begin{figure*}
%\epsscale{1.2}
\plotone{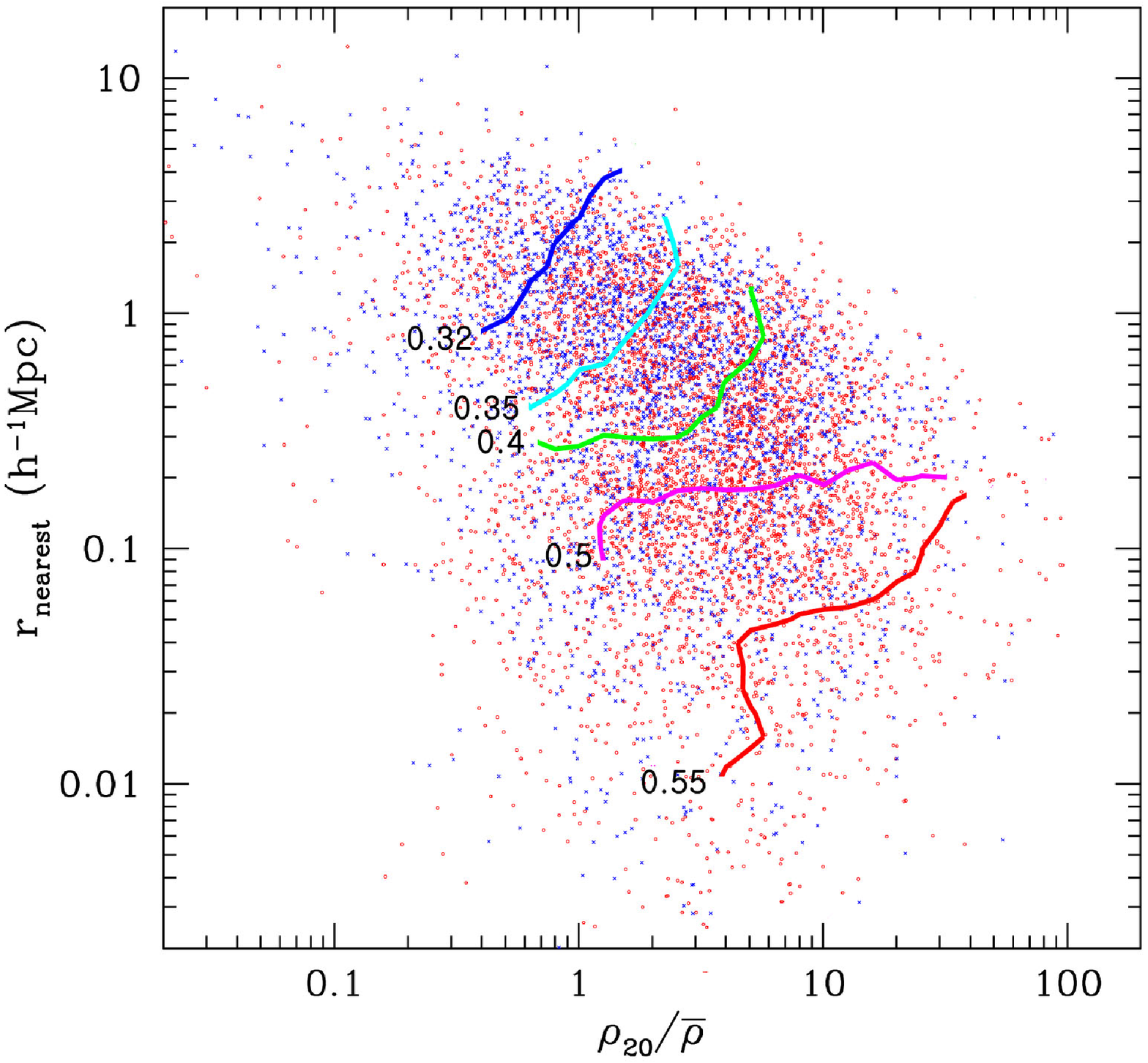}
\caption{Relation between the distance to the nearest companion and
the local density with $N_{s}=20$ smoothing.
Dots are early ({\it red}) and late ({\it blue}) type galaxies 
with $-22.0\leq M_r \leq -20.0$
in the D2 sample. Contours of constant early type galaxy fraction are superposed
on the scatter plot.
}
\end{figure*}
For each galaxy in Sample D2 ($M_{r,\rm lim} = -18.5$) that
has absolute magnitude limits $-20.0\geq M_{r,i} \geq -22.0$
we find the nearest companion that
lies in the range $M_{r,i}+1.5 >M_r >M_{r,i}-2$ 
and has radial velocity difference less 
than 500 km s$^{-1}$, which is a little more generous than that used above.
If luminosity is proportional to mass, 1.5 mag fainter luminosity
corresponds to 25\% of mass.
Our criterion implies that a companion with mass greater than 25\%
of a target galaxy's mass is heavy enough to influence the morphology of the
galaxy.
Superposed are again contours of constant early type fractions of
0.32, 0.35, 0.40, 0.50, and 0.55.
It shows interesting dependences of morphology fraction on both scales.
In general, the contours have positive slopes, which means 
the early type fraction increases as $r_{\rm near}$ decreases
at fixed $\rho_{20}/{\bar{\rho}}$. At fixed $r_{\rm near}$
the fraction increases as $\rho_{20}/{\bar{\rho}}$ increases,
which is already shown in Figure 18 for $r_{\rm near}
= 0.5 h^{-1}$Mpc. The strongest dependence
on $r_{\rm near}$ is observed 
%at intermediate and high background
%densities of $\rho_{20}/{\bar{\rho}}= 1 \sim 20$ and
when $r_{\rm near}\approx 0.2 h^{-1}$ Mpc.
The dependence on $r_{\rm near}$ is weaker when $r_{\rm near}$
is higher or lower.
In under-dense regions below $\rho_{20}/{\bar{\rho}}<0.5$ the majority of 
galaxies are isolated, with $r_{\rm near}>0.5 h^{-1}$Mpc, and
the morphology fraction is much less dependent on the existence of
neighboring galaxies. In this under-dense environment galaxies are
mostly of late type and morphology is sensitive to
$\rho_{20}/{\bar{\rho}}$.
The slope of contours depends on our choice of companion criteria and
the smoothing size used to calculate the morphology fraction.
But the general trend described above remains robust.
A previous study analogous to ours is Balogh et al. (2004b) who claimed
that the fraction of emission line galaxies depends on
both $3.85$ and $0.77 h^{-1}$ Mpc Gaussian smoothing scales.
Importantly, they gave some evidence that the fraction was mostly
dependent on local density $\rho_{0.77}$ where $\rho_{3.85}$ is high,
but depended more on background density $\rho_{3.85}$ where
$\rho_{3.85}$ is lower. Their claim is in agreement with our findings
(compare their Fig. 9 with Fig. 20 of this paper).

%%%%%%%  TABLE 2 %%%%%%%%%%
\begin{deluxetable*}{lccccc}
%\tabletypesize{\footnotesize}
\tabletypesize{\small}
\tablecolumns{6}
\tablewidth{0pt}
\tablecaption{Local Density-dependent Schechter Function Parameters for 
the D3 sample}
\tablehead{
\colhead{Morphology} & 
\colhead{$\rho /\bar{\rho}$}&
\colhead{$\phi^{\ast}$}&
\colhead{$M^{\ast}-5{\rm{log}_{10}}h$}&\colhead{$\alpha$}&
\colhead{${\bar{n}_{\rm gal}}^{~~~~\rm a}$}\\
\colhead{Type}&&
\colhead{$10^{-3} h^3 {\rm Mpc}^{-3}$}&&&
\colhead{$10^{-3} h^3 {\rm Mpc}^{-3}$}
}
\startdata
all& all     &$18.47 $  &$-20.22\pm 0.03$  &$ -0.81\pm 0.03 $  &$ 16.5$\\
  &$ 0.37 $  &$ 3.59 $  &$-19.85\pm 0.07$  &$ -0.70\pm 0.10 $  &$ 2.2 $ \\ 
  &$ 1.39 $  &$ 7.36 $  &$-20.11\pm 0.04$  &$ -0.74\pm 0.05 $  &$ 5.4 $ \\ 
  &$ 4.59 $  &$ 7.18 $  &$-20.21\pm 0.04$  &$ -0.69\pm 0.04 $  &$ 5.5 $ \\ 
  &$14.85 $  &$ 1.50 $  &$-20.61\pm 0.08$  &$ -0.99\pm 0.07 $  &$ 1.7 $ \\ 
\tableline
early& all &$ 7.11$& $ -20.23 \pm0.04$& $ -0.53 \pm 0.04$&$ 5.3$\\
  &$ 0.38 $&$ 0.88$& $ -19.95 \pm0.13$& $ -0.56 \pm 0.16$&$ 0.6$ \\
  &$ 1.43 $&$ 2.65$& $ -19.98 \pm0.06$& $ -0.13 \pm 0.08$&$ 1.7$ \\
  &$ 4.75 $&$ 2.92$& $ -20.24 \pm0.06$& $ -0.47 \pm 0.07$&$ 2.2$ \\
  &$15.53 $&$ 0.79$& $ -20.68 \pm0.11$& $ -0.87 \pm 0.10$&$ 0.9$ \\
\tableline
late& all  &$12.27$&$  -20.12  \pm  0.03$&$ -0.90 \pm  0.04 $&$ 9.4$\\
  &$ 0.37 $&$2.76 $&$  -19.80 \pm  0.09$&$ -0.75 \pm  0.11 $&$ 1.6 $ \\
  &$ 1.39 $&$4.67 $&$  -20.14 \pm  0.06$&$ -0.93 \pm  0.06 $&$ 3.6 $ \\
  &$ 4.59 $&$4.54 $&$  -20.10 \pm  0.06$&$ -0.77 \pm  0.06 $&$ 3.3 $ \\
  &$14.37 $&$0.75 $&$  -20.51 \pm  0.14$&$ -1.09 \pm  0.12 $&$ 0.8 $ \\
\enddata
\tablenotetext{a}{A fraction of 0.505 of late type galaxies with $b/a > 0.6$,
is used to infer the total number density of late type galaxies.}
%\tablecomments{}
\end{deluxetable*}

% MSV: Do we need a discussion section to relate this results to other
% papers on environmental dependence (e.g., Rojas et al., Blanton et al., 
% and many other refs)?

\section{Discussion}

\subsection{Summary}

We have studied dependences of physical properties of galaxies in the
Sloan Digital Sky Survey (SDSS) on environment. There are several
differences between our analysis and previous studies. First, we have
used smooth local density estimated by the spline kernel enclosing 20
or 200 $L_\ast$ galaxies to define environment. Because the kernel used
is centrally weighted, finite, and variable in size, it gives us a
wide dynamic range in the estimated local density over 4 orders of
magnitude.  Second, we have used an accurate automated morphology
classifier to group large numbers of galaxies into early and late
types. In the case of the D2 sample with minor correction by visual
inspection we expect the completeness and reliability to exceed
93\%. These high quality morphology sets significantly reduce the
biases due to misclassification. Third, we have used a set of
volume-limited samples defined by absolute magnitude. Environmental
dependences are studied for these volume-limited samples at fixed
morphology and luminosity. In this way we have avoided confusion due
to mixed dependences that occur when flux-limited samples are used.
We have also attempted to reduce the effects due to internal
extinction by excluding late type galaxies with axis ratios less than
$b/a=0.6$.  Thanks to the large size of the SDSS spectroscopic catalog
of galaxies, we are able to measure the environmental dependences to
reasonable accuracies even after dividing galaxies into many
subsamples selected by morphology and luminosity. Our major findings
are as follows.

1. The fraction of early morphological type galaxies is a
monotonically increasing function of a (well-defined) local density 
and luminosity. 
%At
%a given luminosity and local density the early-to-late
%fraction is fixed. 
% MSV: ``Fixed'' might be too strong a word. There is some scatter
% in this relation from one patch of space to another.
The morphology-density-luminosity
relation, as measured in this work, should be a key constraint on
galaxy formation models.
It remains for us to understand what causes the morphology of a galaxy
at fixed local density. We find that morphological fractions vary
strongly with local density, but we find a mix of both early and late
types at all densities. Thus, conditions other than local density must
contribute to determining the morphology of a galaxy at fixed local
density.
The extra cause might be environmental properties other than density.
The gravitational potential shear field is an example of such
an environmental property.
Morphology can also depend on the merger history of each object,
which can be stochastic to some degree.
%A number of recent investigations (e.g. Bernardi et al. 2003b;
%Hogg et al. 2004; Kauffmann et al. 2004; Balogh et al. 2004a,b; 
%Quintero et al. 2006) have indicated that
%while the fraction of early type galaxies increases with local
%density, the scaling relations of early- and late type galaxies are
%left relatively unchanged by environment. Our results have refined the
%previous work by classifying galaxies in a way more closely related to
%their visual morphology. We find that once galaxies are classified
%this way, there is very little residual dependence of their properties
%on environment. Thus, if galaxies transform from late type to
%early type during their evolution, {\it how often} they do so depends
%on the local environment but {\it how} they do so (or at least their
%final state) does not.

2. We find that the environmental dependences of various galaxy
properties are almost entirely due to the dependences of morphology
and luminosity on environment. When morphology and luminosity are
fixed, other physical properties, such as color gradient, concentration, size,
velocity dispersion, $H{\alpha}$ emission, and axis ratio, are nearly
independent of local density. Careful inspection reveals weak residual
local density dependences for several properties.  We find that the
$u-r$ color of late type galaxies is monotonically bluer at low
density environments, even at a fixed luminosity.  Early types show
much weaker dependence.  We also find that bright early type galaxies
in high density regions have higher central velocity dispersions
compared to those in low density regions, but that the opposite is
true for fainter galaxies.  A consequence is a local density
dependence of the $L$-$\sigma$ relation.  
Because galaxy properties
strongly depend not only on luminosity but also on morphology, it is
clear that galaxy properties cannot be determined solely by dark halo
mass. 

% MSV: I'm not sure what you mean by the following sentence.
% This paper shows that, at fixed luminosity, the E fraction varies with
% environment, so it seems clear that local density affects morphology.
%There must be another important cause for the morphology of
%a galaxy in a dark halo of a given mass. 

Our results, using a new morphological classifier and a robust
method of local density estimation, are largely consistent with 
several previous studies attempts
(Kauffmann et al. 2004;
Blanton et al. 2005a; Hashimoto \& Oemler 1999; Christlein \& Zabludoff 2005).
In each of those papers,
the authors found a relationship between galaxy color and environment
at fixed concentration, a measure of galaxy structure 
associated with morphology (Strateva et al. 2001).  Conversely, those
papers showed that at fixed color there is no relationship between
concentration and environment.  Those results regarding concentration
do not refer to morphology as we define it here, which indeed includes
$u-r$ color as one of key indicators. In Park \& Choi (2005) we have shown 
that morphology can be traced not simply by concentration alone, 
but by a combination of concentration, color, and color gradient.
Once morphology is defined this way, almost any
other indicator of type appears to be nearly fixed with environment,
even as the fraction of galaxies of each morphology changes. 
%Thus, our
%results are perfectly consistent with the previous ones. 
Indeed, they
imply that because the morphological mix changes with local density,
at fixed concentration one should indeed measure a correlation between
color and density.  The converse result, that concentration shows only
a weak relationship with density at fixed color, is a consequence of
the fact that color is the more powerful type indicator.

3. We show that morphology of galaxies appears to depend on
local density, from very small scales ($\leq 0.1 h^{-1}$ Mpc)
up to the $N_s =20$ and 200 smoothing scales corresponding to about 4.7 and
$12 h^{-1}$ Mpc effective Gaussian smoothing scales.
It is shown that the environment at scales larger than about
$12 h^{-1}$ Mpc do not independently affect galaxy morphology.
This gives us an important clue to understanding the origin of galaxy
morphology.
The direct cause of galaxy morphology is not related with the
background density of the linear-regime scale at the present epoch.
We have also shown that morphology of bright galaxies 
depends on both the local density with
$N_s =20$ smoothing and the distance to the nearest companion galaxy.
%This is particularly so in under-dense regions.
The dependence on $r_{\rm near}$ becomes strongest when the
neighbor distance is about $200 h^{-1}$ kpc at which the combined
effects of strong tidal force and long duration of interaction 
on galaxy property seems to reach the maximum.
The importance of the smooth density with $N_s =20$ rises at
larger and smaller separations.

\subsection{How Is the Galaxy Morphology Determined?}

In the currently popular scenario of galaxy formation and evolution
the properties of galaxies depend either only on the formation history
of their dark matter halos or also on nurture processes such as ram pressure
stripping, strangulation, and harassment.
In the cold dark matter (CDM) cosmogony, a galaxy-scale dark halo forms through a series
of mergers of subgalactic objects. The over-density at the galaxy scale,
which is statistically related with the larger scale density,
will determine the final mass.
Other physical properties of the galaxy will depend on the final mass,
but much variation is expected, depending on
how the final mass is actually assembled.
Gott \& Thuan (1976) proposed that galaxy morphology is
determined by the amount of gas left over at maximum collapse
of the protogalaxy, which in turn depends on the ratio of the
star formation time scale $\tau_{\rm SF}$ to the collapse time $\tau_{\rm coll}$.
They claimed that ellipticals had relatively small $\tau_{\rm SF}/\tau_{\rm coll}$
(if a $\rho^2$ star formation law was adopted) and had finished
star formation by the time of maximum collapse.
In additional to this initial condition and sub-galactic-scale evolution,
the local environment can also directly affect galaxy properties.
Ram pressure stripping (Gunn \& Gott 1972)
is a mechanism that removes gas and shuts off star formation when a galaxy
orbits a hot cluster halo. The morphology of a spiral galaxy can be
transformed to an early type by this process.
High-speed encounters of galaxies with other halos typically in clusters
cause impulsive heatings, called harassment (Moore et al. 1996),
and can transform spirals to early types.
Another mechanism, strangulation (Balogh et al. 2000),
can also transform morphology through a decline of star formation rate
due to shut-off of the newly accreted gas when a galaxy enters
a cluster or group environment and loses its hot gas reservoir.
There is also observational and theoretical evidence that the tidal
force field in clusters can transform the infalling spirals to early
types (Moss \& Whittle 2000; Gnedin 2003).

It is important to note that all these nurture mechanisms proposed
so far are basically effective in group or cluster environments.
In previous studies it was claimed that galaxy properties such as star
formation activity and morphology fraction showed
discontinuities at a critical density for galaxies somewhat
fainter than the $L_\ast$ galaxy. This finding was used to support
the proposition that properties of the relatively faint galaxies are affected by
morphology transformation mechanisms working in the group or
cluster environment (Tanaka et al. 2004; Balogh et al. 2004).
However, we find that
there is no feature in the galaxy property versus local density relation
even for such relatively faint galaxies when luminosity and morphology are fixed.
The only feature is the sharp decrease of the late-type fraction above
the critical luminosity of about $M_r = -21.3$ in the morphology
versus luminosity relation (Paper I).
This indicates that there seems to be morphology and luminosity-determining
processes that work at all local densities in a continuous way.
%and that the cluster or group environment has no special role in determining
%the morphology and other properties of galaxies and

The strong effect of a close companion on galaxy morphology
as indicated by this study, might imply the existence of a morphology
transformation mechanism acting at a distance without a
direct physical contact.  What is important is that this process
seems to be working in a wide density environment.
Figure 20 indicates that this effect is maximal
for galaxies having a companion at a separation of about $200h^{-1}$ kpc.
This might be because the tidal effects are strong enough to significantly
enhance the star formation rate, and the duration of interaction is large
as well at this particular separation.
The morphology transformation process caused by this rather secular
effects in binary or few-body systems can be smoothly connected
with the more instantaneous tidal interactions with the cluster core
or local substructures in clusters 
(Moss \& Whittle 2000; Gnedin 2003).

Galaxies can be either isolated or in binary/multiple systems
at all environments.  For example, in Figure 18 the fraction of
unisolated galaxies is about 20\% even in the low density region with
$\rho_{20}/{\bar \rho}\approx 0.5$.  A similar finding was reported
in the existence of red galaxies in low density regions (Balogh et al. 2004b).
Figure 18 also shows that the early type fraction is systematically higher
for galaxies located at higher $\rho_{20}/{\bar \rho}$ no matter whether
they are isolated or in systems.
The isolated early types may have been initially born as early types or
may be final merger products of initially multiple systems.
%Since both initial early type fraction and average merger rate will be
%increasing function the local background density, the systematic
%dependence of the early type fraction on $\rho_{20}/{\bar \rho}$ can be
%explained.

It is reasonable to believe that our local background density
$\rho_{20}$ calculated by using neighboring 20 $L_\ast$ galaxies is a
good indicator of the ratio $\tau_{\rm SF}/\tau_{\rm coll}$ at the time of galaxy
formation. The merger rate during the subsequent evolution and
the average rates of all the nurture processes mentioned above
are likely to be increasing functions of $\rho_{20}$. The fraction of galaxies
in binary/multiple systems are also an increasing function of $\rho_{20}$.
If these are true and if small $\tau_{\rm SF}/\tau_{\rm coll}$ or above evolution processes 
yield early type galaxies, we can understand the overall dependence
of morphology on $\rho_{20}$.
The increase of the early type fraction in binary or multiple systems
can be explained by the tidal disturbance effects.
It remains to be understood why the dependence on the nearest companion
distance becomes maximum near the particular separation of $200 h^{-1}$ kpc,
which is roughly the separation between two typical bright galaxies
when their dark halos are just touching.

\acknowledgments
The authors thank Chan-Gyung Park for measuring
the luminosity function and Yeong-Shang Loh for helpful comments.
CBP acknowledges the support of the Korea Science and Engineering
Foundation (KOSEF) through the Astrophysical Research Center for the
Structure and Evolution of the Cosmos (ARCSEC).
% and through the grant R01-2004-000-10520-0. 
MSV acknowledges support from NASA grant
NAG-12243 and NSF grant AST-0507463. MSV thanks the Department of
Astrophysical Sciences at Princeton University for its hospitality
during sabbatical leave. YYC, CBP, and MSV thank the Aspen Center for
Physics, at which much of this paper was written. JRG is supported by
NSF GRANT AST 04-06713.

Funding for the SDSS and SDSS-II has been provided by the Alfred P. Sloan 
Foundation, the Participating Institutions, the National Science 
Foundation, the U.S. Department of Energy, the National Aeronautics and 
Space Administration, the Japanese Monbukagakusho, the Max Planck 
Society, and the Higher Education Funding Council for England. 
The SDSS Web Site is http://www.sdss.org/.

The SDSS is managed by the Astrophysical Research Consortium for the 
Participating Institutions. The Participating Institutions are the 
American Museum of Natural History, Astrophysical Institute Potsdam, 
University of Basel, Cambridge University, Case Western Reserve University, 
University of Chicago, Drexel University, Fermilab, the Institute for 
Advanced Study, the Japan Participation Group, Johns Hopkins University, 
the Joint Institute for Nuclear Astrophysics, the Kavli Institute for 
Particle Astrophysics and Cosmology, the Korean Scientist Group, the 
Chinese Academy of Sciences (LAMOST), Los Alamos National Laboratory, 
the Max-Planck-Institute for Astronomy (MPIA), the Max-Planck-Institute 
for Astrophysics (MPA), New Mexico State University, Ohio State University, 
University of Pittsburgh, University of Portsmouth, Princeton University,
the United States Naval Observatory, and the University of Washington. 
{}

\begin{thebibliography}{}
\bibitem[],{} Abazajian, K., et~al. 2004, \aj, 128, 502
\bibitem[],{} Adelman-McCarthy, J.~K., et~al. 2007, \apjs submitted
\bibitem[],{} Andreon, S., \& Cuillandre, J.-C. 2002, \apj, 569, 144
\bibitem[],{} Balogh, M., et~al. 2004a, \mnras, 348, 1355
\bibitem[],{} Balogh, M.~L., Baldry, I.~K., Nichol, R., Miller, C., Bower, R., \&
  Glazebrook, K. 2004b, \apjl, 615, L101
\bibitem[],{} Bernardi, M., Nichol, R.~C., Sheth, R.~K., Miller, C.~J., \&
  Brinkmann, J. 2006, \aj, 131, 1288
\bibitem[],{} Bernardi, M., et~al. 2003a, \aj, 125, 1817
\bibitem[],{} Bernardi, M., et~al. 2003b, \aj, 125, 1866
\bibitem[],{} Blanton, M.~R., \& Berlind, A.~A. 2007, \apj in press 
(astro-ph/0608353)
\bibitem[],{} Blanton, M.~R., Eisenstein, D., Hogg, D.~W., Schlegel, D.~J., \&
  Brinkmann, J. 2005a, \apj, 629, 143
\bibitem[],{} Blanton, M.~R., et~al. 2005b, \apj, 129, 2562 % vagc
\bibitem[],{} Blanton, M.~R., Lin, H., Lupton, R.~H., Maley, F.~M., Young, N., Zehavi, I., \& Loveday, J. 2003a, \aj, 125, 2276 % tiling
\bibitem[],{} Blanton, M.~R., et~al. 2003b, \aj, 125, 2348 % K
\bibitem[],{} Blanton, M.~R., et~al. 2003c, \apj, 594, 186 % properties
%\bibitem[],{} Butcher, H.~R., \& Oemler, A. 1984, \apj, 285, 426
\bibitem[],{} Boselli, A., \& Gavazzi, G. 2006, \pasp, 118, 517
\bibitem[],{} Choi, Y.-Y., Park, C., \& Vogeley, M.~S. 2006, \apj, submitted (Paper I)
\bibitem[],{} Christlein, D., \& Zabludoff, A.~I. 2005, \apj, 621, 201
\bibitem[],{} Colless, M., et al. 2001, \mnras, 328, 1039 % 2dFRGS
\bibitem[],{} Croton, D. J., et~al. 2005, \mnras, 356, 1155 % 2dFGRS
\bibitem[],{} de Propris, R., Pritchet, C.~J., Harris, W.~E., \& McClure, R.~D. 1995,
  \apj, 450, 534
\bibitem[],{} De Propris, R., et~al. 2003, \mnras, 342, 725
%\bibitem[],{} De Propris, R., et~al. 2004, \mnras, 351, 125
\bibitem[],{} Dressler, A. 1980, \apj, 236, 351
\bibitem[],{} Driver, S.~P., Couch, W.~J., \& Phillipps, S. 1998, \mnras, 301, 369
%\bibitem[],{} Eisenstein, D. J., et~al. 2001, \aj, 122, 2267
%\bibitem[],{} Ellingson, E., Lin, H., Yee, H.~K.~C., \& Carlberg, R.~G. 2001, \apj,
%  547, 609
\bibitem[],{} Fukugita, M., Ichikawa, T., Gunn, J.~E., Doi, M., Shimasaku, K., \& Schneider, D.~P. 1996, \aj, 111, 1748
\bibitem[],{} Gnedin, O., Y. 2003, \apj, 589, 752
\bibitem[],{} G\'{o}mez, P., et~al. 2003, \apj, 584, 210
\bibitem[],{} Goto, T., et~al. 2003, \mnras, 346, 601
\bibitem[],{} Gott, J.~R., \& Thuan, T.~X. 1976, \apj, 204, 649
\bibitem[],{} Gunn, J.~E., et~al. 2006, \aj, 131, 2332 % 2.5m telescope
\bibitem[],{} Gunn, J.~E., et~al. 1998, \aj, 116, 3040
\bibitem[],{} Gunn, J.~E., \& Gott, J.~R. 1972, \apj, 176, 1
\bibitem[],{} Hashimoto, Y. \& Oemler, A.~J. 1999, \apj, 510, 609
\bibitem[],{} Hamilton, A.~J.~S. \& Tegmark, M. 2004, \mnras, 349, 115
\bibitem[],{} Hogg, D.~W., et~al. 2004, \apjl, 601, L29
\bibitem[],{} Hogg, D.~W., Finkbeiner, D.~P., Schlegel, D.~J., \& Gunn, J.~E. 2001, \aj, 122, 2129
\bibitem[],{} Hoyle, F., Rojas, R.~R., Vogeley, M.~S., \& Brinkmann, J. 2005, \apj,
  620, 618
\bibitem[],{} Ivezi\'{c}, Z., et~al. 2004, AN, 325, 583
\bibitem[],{} J{\o}rgensen, I., Franx, M., \& Kj{\ae}rgaard, 1995, \mnras, 276, 1341
\bibitem[],{} Kauffmann, G., White, S.~D.~M., Heckman, T.~M., M\' enard, B.,
  Brinchmann, J., Charlot, S., Tremonti, C., \& Brinkmann, J. 2004,
  \mnras, 314
\bibitem[],{} Lewis, I., et~al. 2002, \mnras, 334, 673
\bibitem[],{} Lumsden, S.~L., Collins, C.~A., Nichol, R.~C., Eke, V.~R., \& Guzzo,
  L. 1997, \mnras, 290, 119
\bibitem[],{} Lupton, R.~H., Gunn, J.~E., Ivez\'{i}c, Z., Knapp, G.~R., Kent, S., \& Yasuda, N. 2001, in ASP Conf. Ser. 238, Astronomical Data Analysis Software and Systems X, ed. F.~R. Harnden, Jr., F.~A. Primini, \& H.~E. Payne (San Francisco: ASP), 269
%\bibitem[],{} Margoniner, V.~E., de Carvalho, R.~R., Gal, R.~R., \& Djorgovski, S.~G.
%  2001, \apjl, 548, L143
\bibitem[],{} Marinoni, C., Hudson, M.~J., \& Giuricin, G. 2002, \apj, 569, 91
\bibitem[],{} Moss, C., Whittle, M. 2000, \mnras, 317, 667
\bibitem[],{} Oemler, A. 1974, \apj, 194, 1
\bibitem[],{} Park, C., Choi, Y.-Y., \& Kim, J. 2006, in preparation
\bibitem[],{} Park, C., et~al. 2005, \apj, 633, 11
\bibitem[],{} Park, C., \& Choi, Y.-Y., 2005, \apj, 635, L29
\bibitem[],{} Park, C., Vogeley, M.~S., Geller, J., \& Huchra, J.~P. 1994, \apj, 431,569
\bibitem[],{} Pier, J.~R., Munn, J.~A., Hindsley, R.~B., Hennessy, G.~S., Kent, Si.~M., Lupton, R.~H., \& Ivez\'{i}c, R. 2003, \aj, 125, 1559
\bibitem[],{} Popesso, P., B\" ohringer, H., Romaniello, M., \& Voges, W. 2005,
  \aap, 433, 415
\bibitem[],{} Popesso, P., Biviano, A., B\"ohringer, H., \& Romaniello, M. 2006,
  \aap, 445, 29
\bibitem[],{} Postman, M., \& Geller, M. 1984, \apj, 281, 95
\bibitem[],{} Quintero, A.~D., Berlind, A.~A., Blanton, M.~R., \& Hogg, D.~W. 2006, \apj,
  submitted (astro-ph/0512004)
\bibitem[],{} Rojas, R. R., Vogeley, M. S., Hoyle, F., \& Brinkmann, J. 2004, \apj, 617, 50
\bibitem[],{} Rojas, R. R., Vogeley, M. S., Hoyle, F., \& Brinkmann, J. 2005, \apj, 624, 571
%\bibitem[],{} Richards, G.~T., et~al. 2002, \aj, 123, 2945
\bibitem[],{} Schlegel, D.~J., Finkbeiner, D.~P., \& Davis, M. 1998, \apj, 500, 525
\bibitem[],{} Sandage, A., \& Visvanathan, N. 1978, \apj, 225, 742
\bibitem[],{} Secker, J., Harris, W.~E., \& Plummer, J.~D. 1997, \pasp, 109, 1377
%\bibitem[],{} Smail, I., Edge, A.~C., Ellis, R.~S., \& Blandford, R.~D. 1998, \mnras,
%  293, 124
\bibitem[],{} Smith, J.~A., et~al. 2002, \aj, 123, 2121
\bibitem[],{} Stoughton, C., et~al. 2002, \aj, 123, 485
\bibitem[],{} Strateva, I., et~al. 2001, \aj, 122, 1861
\bibitem[],{} Strauss, M.~A., et~al. 2002, \aj, 124, 1810
\bibitem[],{} Tanaka, M., et~al. 2004, \aj, 128, 2677
\bibitem[],{} Tegmark, M., et~al. 2004, \apj, 606,702
\bibitem[],{} Trentham, N. 1998, \mnras, 293, 71
\bibitem[],{} Thuan, T.~X., \& Gott, J.~R. 1977, \apj, 216, 194
\bibitem[],{} Tucker, D., et~al. 2006, AN, in  press
\bibitem[],{} Valotto, C.~A., Moore, B., \& Lambas, D.~G. 2001, \apj, 546, 157
\bibitem[],{} Valotto, C.~A., Nicotra, M.~A., Muriel, H., \& Lambas, D.~G. 1997,
  \apj, 479, 90
\bibitem[],{} Weinmann, S.~M., van den Bosch, F.~C., Yang, X., \& Mo, H.~J. 2006, \mnras, 366, 2
\bibitem[],{} York, D., et~al. 2000, \aj, 120, 1579
\bibitem[],{} Zandivarez, A., Mart\'{i}nez, H.~J., \& Merch\'{a}n, M.~E. 2006, astro-ph/0602405
%\bibitem[],{} Zehavi, I. et~al. 2002, \apj, 571, 172
\end{thebibliography}
\end{document}